\documentclass[letterpaper]{article}
\usepackage{aaai20}
\usepackage{times}
\usepackage{helvet}
\usepackage{courier}
\usepackage[hyphens]{url} 
\usepackage{graphicx}
\usepackage{multirow}
\usepackage{subcaption}
\usepackage{booktabs}
\usepackage{xcolor}
\usepackage{appendix}
\usepackage{array}
\usepackage{tabularx}
\usepackage{graphicx}
\newcommand{\citet}[1]{\citeauthor{#1} \shortcite{#1}}
\newcommand{\citep}{\cite}

\title{Ginger Cannot Cure Cancer: Battling Fake Health News with a Comprehensive Data Repository}

\author{Enyan Dai, Yiwei Sun, Suhang Wang\\
The Pennsylvania State University\\
\{emd5759, yus162, szw494\}@psu.edu\\
}

\begin{document}
\maketitle
\begin{abstract}


Nowadays, Internet is a primary source of attaining health information. Massive fake health news which is spreading over the Internet, has become a severe threat to public health. Numerous studies and research works have been done in fake news detection domain, however, few of them are designed to cope with the challenges in health news. For instance, the development of explainable is required for fake health news detection.  
To mitigate these problems, we construct a comprehensive repository, FakeHealth, which includes news contents with rich features, news reviews with detailed explanations, social engagements and a user-user social network. Moreover, exploratory analyses are conducted to understand the characteristics of the datasets, analyze useful patterns and validate the quality of the datasets for health fake news detection. We also discuss the novel and potential future research directions for the health fake news detection.
\end{abstract}
\section{Introduction}
The online health information has become an important source for medical advice.
According to \citet{finney2019online}, 81.5\% of U.S. population searchs for health or medical information online, and 68.9\% of the U.S. adults consider the internet first to seek health information. 
However, the health information has been contaminated by various disinformation \cite{schwitzer2017pollution}. Furthermore, with the advent of social media platforms, the fake news pieces become easier to access and more influential. 
For example, a fake health news piece debunked by scientists, ``Ginger is 10,000x more effective at killing cancer than chemo'', generated around 1 million engagements on Facebook\footnote{https://www.nbcnews.com/news/us-news/social-media-hosted-lot-fake-health-news-year-here-s-n1107466}. 
The flood of false medical news is threatening the public health. For instance, a cancer patient mistook an online ad for an experimental cancer treatment as medically reliable information, which resulted in his death\footnote{https://www.bbc.com/news/business-36189252}.
Therefore, fake health news detection is a critical problem which requires more attention.

Fake health news detection has several challenges.
\textit{First}, sophisticated fake health news could disguise itself with the tricks that are hard to be noticed. For example, it may mislead the readers by stating the association as causation or mixing up the absolute risk and relative risk\footnote{https://www.healthnewsreview.org/toolkit/tips-for-understanding-studies/}, which only require minor modifications based on the true information. 
\textit{Second}, compared with the news in political, sports and shopping domain, social media users may be less likely to make accurate judgements on the credibility of health news. Because the identification of fake news in health care tends to require specialist knowledge.
This could bring challenges to the fake news detection approaches based on social media user opinions.
\textit{Third}, explainable health fake news detection is crucial because: (i) explanations can make social media users trust the prediction results and stop propagating fake health news; and (ii) the explanations help practitioners to understand if the output of the classifiers are as expected and if there's sample bias on the training dataset. 
Whereas, the existing fake news detection methods are opaque especially the deep learning models.

Despite the urgent need to develop novel algorithms to identify fake health news, few work focused on this topic. One of the main reasons is the lack of comprehensive fake health news dataset. 
To address the problems of fake health news detection, we build a comprehensive repository named \textit{FakeHealth}\footnote{https://doi.org/10.5281/zenodo.3606757}, which consists of two datasets, i.e., \textit{HealthStory} and \textit{HealthRelease}. Each dataset includes news contents, news reviews, social engagements and user networks.
For the news contents, text is provided along with the source publishers, image links and other side information, 
which benefits the development of algorithms utilizing publisher credibility, visuals and other related information in addition to article norm or style for better detection performance.
Besides, source news social engagements are supplemented with user profiles, user timelines and friends profiles. The abundant information of users could help us develop sophisticated models which pay more attention to the users that are less likely to be misled.
In the end, we obtained 500k tweets, 29k replies, 14k retweets, 27k users profiles with timelines and friend lists.  What's more, the news reviews cover explanations regarding ten health news evaluation criteria. These explanations point out the aspects that the health news presents poorly, which allows us to develop the explainable fake health news detection approaches by recognizing the criterion the health news fails to meet. Furthermore, the background knowledge described in the reviews could be applied to build health knowledge graph as interpretable model. The main contributions of this paper are: 
\begin{itemize}
    \item We construct and release the first comprehensive fake health news datasets. The rich features of our repository including news contents, news reviews and social contexts make it possible to detect fake health news with various approaches for better performance and interpretability.   
    \item We perform some exploratory analyses on the datasets to validate the quality of the datasets and to understand the key properties of the datasets.
    \item We discuss potential and urgent research directions such as explainable fake news detection and knowledge-based fake news detection that can be conducted on our datasets.
\end{itemize}

\section{Related Work}

Fake news detection has attracted increasing attention and many fake news detection datasets are developed and publicly available. Most of them only contain news contents.
For example, \textbf{BuzzFeedNews}\footnote{https//github.com/BuzzFeedNews/2016-10-facebook-fact-check/} specializes in political news published on Facebook during the 2016 U.S. Presidential Election. The dataset contains 1627 news articles checked by 5 Buzzfeed journalists. They give a binary label to per article with the news source. 
\textbf{LIAR} \cite{wang2017liar} has 12.8k short statements with manual labels from the political fact-checking website. It is useful to find misinformation in short statements, but can't be applied for complete news articles.
\textbf{NELA-GT-2018} \cite{norregaard2019nela} provides 714k news articles in general topics from 194 news producers. The labels are obtained from 8 assessment sites. Among the 194 news sources, 40 of them are not found any labels from the assessments sites. 
\textbf{FA-KES} \cite{salem2019fa} is a fake news dataset around Syrian war. It consists of 804 articles labeled as real or fake. The labels were annotated based on the database from Syrian Violations Documentation Center with a cluster algorithm, so the reliability of these labels may bring concerns. Apart from their own limitations, the common drawback of the data listed here is the short of social context and the information other than the text of news articles. 

In addition to news contents, several datasets also contain social context such as user comments and reposts on the social media platforms. 
\textbf{CREDBANK} \cite{mitra2015credbank} contains about 1000 news events whose credibility labeled by Amazon mechanical Turk. There are 60 million tweets between 2015 and 2016 in the dataset, but the original news articles of the events are not included. 
\citet{ma2016detecting} collected \textbf{Twitter} and \textbf{Weibo} for their fake news detection approach. For the \textbf{Twitter}, they obtained 778 reported events from March 2015 to December 2015 which received 1 million posts from 500k users. And they confirmed rumors and non-rumors from www.snopes.com as labels. As for \textbf{Weibo}, it consists of 2313 rumors which are reported by Sina community management center. The 2351 non-rumors are crawled from the general threads which are not reported as rumors. Around 4 million posts from 2 million users are acquired for the rumors and non-rumors in total.   
 \textbf{FacebookHoax} \cite{tacchini2017some} contains around 15K Facebook posts about science news. The labels for "hoax" or "non-hoax" are determined by the sources of the Facebook pages. This dataset aims to deal with individual posts instead of news articles.
\textbf{FakeNewsNet} \cite{shu2018fakenewsnet} is a data repository with news content, user engagements, user network and spatiotemporal information for fake news studies. It contains political news which are checked by politifact and gossiocop. Although these engagement-driven datasets are valuable for fake news detection, they mostly haven't cover any user profiles except FakeNewsNet.

While these datasets are useful, they can't address the challenges presented in fake health news. The existing fake news datasets don't cover enough health news contents, barely include user network information and provide no related knowledge and explanations to the ground truth.  Thus, we collect two datasets containing news content with rich features, social engagements with user network and news reviews providing answers for standard health news evaluation criteria to address the issues in fake health news detection.

\section{Dataset Collection}
In this section, we first introduce the HealthNewsReview.org, which provides news reviews with the ground truth labels and corresponding explanations. Then we give the details of the data collection process.
\subsection{HealthNewsReview}
HealthNewsReview.org \footnote{https://www.healthnewsreview.org/} is a web-based project running from 2005 to 2018. It critically analyzes the claims about health care interventions to improve the quality of health care information. HealthNewsReview.org is \textit{free from industry} and supported by Informed Medical Decisions Foundation and Laura and John Arnold Foundation. It runs without accepting any advertising and funding from any entity in conflict of interest. HealthNewsReview.org reviews news stories from main US media and news releases from institutes. The contents include the claims of efficacy about specific treatments, tests, products or procedures. The news pieces are assessed based on a standard rating system. Each news is reviewed by at least two reviewers with years of experience in health domain. The reviewers are from journalism, medicine, health services research, public health and patient, and each of them signs an industry-independent disclosure agreement. The diversity and independence of the reviewers could reduce the bias of the assessments.

\subsection{Data Collection Pipeline}

We collected datasets with the following four steps:
\begin{enumerate}
    \item We crawled the reviews of news stories and releases from HealthNewsReview.org. The source news titles and URLs are included in the collected files.
    \item With the source news titles and URLs, we scraped the news contents which correspond to the acquired reviews.
    \item We obtained the social engagements on Twitter by collecting tweets, replies and retweets about the source news. 
    \item We further supplemented the social context with the engaged user network information. 
\end{enumerate}
\begin{figure}[t]
    \centering
    \includegraphics[width=1\columnwidth]{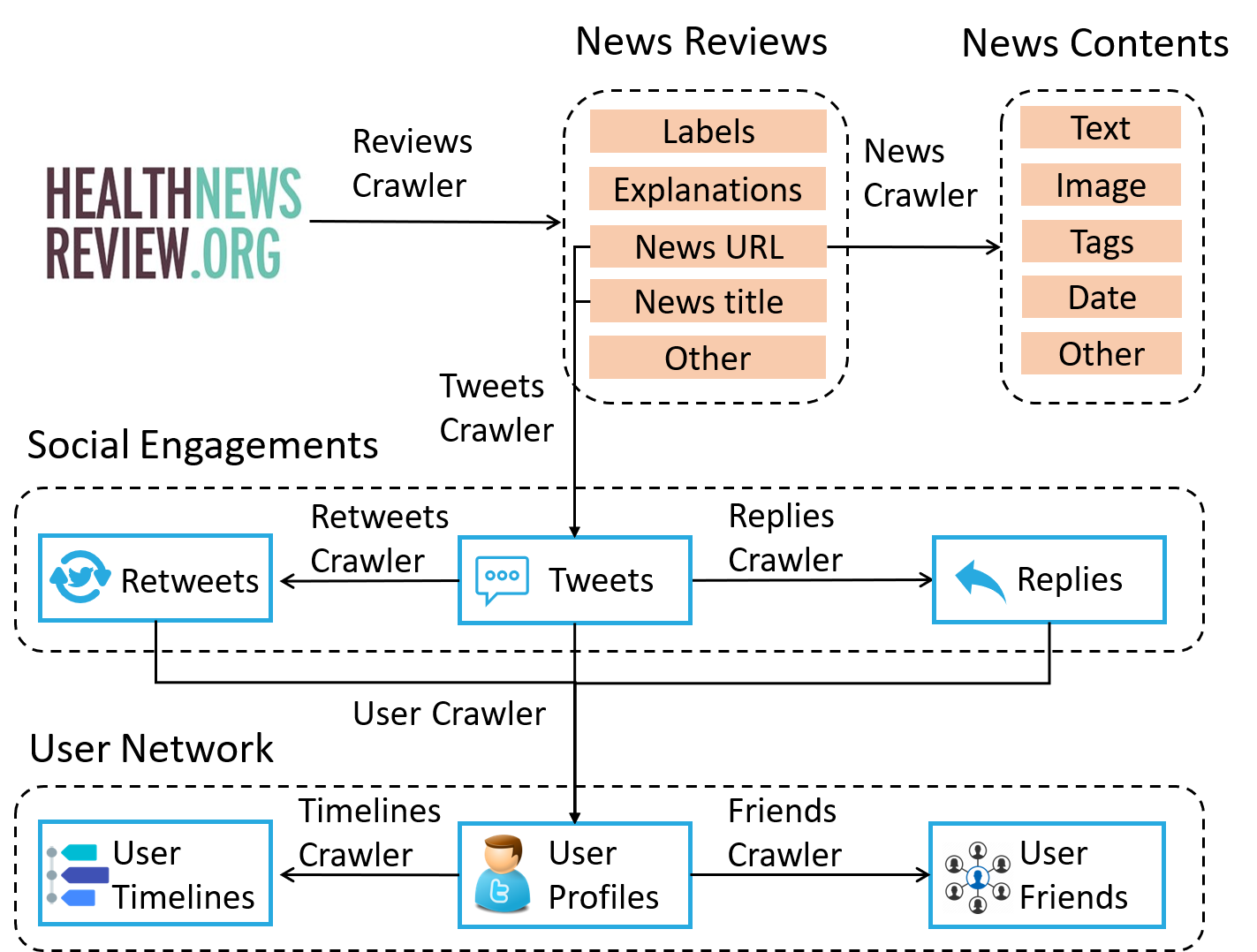}
    \caption{Overview of the collection pipeline.}
    \label{fig:collection_pipeline}
    \vspace{-3mm}
\end{figure}

The collection pipeline is shown in Figure \ref{fig:collection_pipeline}. The description of storage format and the crawling API is presented in the Data Format section of Appendix. Next, we introduce each crawling component in detail.

\textbf{News Reviews and Contents:}  The reviews were crawled complying to the robots.txt of HealthNewsReview.org. In order to ensure that the obtained news have adequate social engagements on Twitter, we filtered out the news reviews published earlier than December 2009, which is a time point that users began to actively interact with the news on Twitter. The news contents are collected with the news source URLs presented on the HewNewsReveiws.org.
In some cases, the source URLs are missing or out of date. In this situation, we checked the archive website Wayback Machine\footnote{https://archive.org/web/} to restore the unavailable links.  

\textbf{Social Engagements:} 
The social context of the source news was attained from the Twitter, which is one of the most popular social media platforms. Twitter provides convenient API to collect user engagements. We adopted Twitter's Advanced Search API\footnote{https//twitter.com/search-advanced?lang=en} to search the tweets which directly post the news stories and releases. In order to collect adequate tweets and restrict the noisy tweets, we searched and supplemented the related tweets by three steps. Firstly, we used the titles as search queries to get tweets. Then, we extracted the key words of the title by removing special tokens and common stop words in the title. In this step, we kept at least five key words to avoid over general queries. Finally, we set the long URL of the news as query to search tweets.
The long links in some tweets may be warped to short URLs by Twitter. Fortunately, the advanced search API will also return the tweets with the short URLs redirecting to the long URLs. This enables we won't miss the tweets disseminated in a shortened form.
With these three strategies, we obtained around 500,000 tweets in total.  After we acquired the tweets, we further scraped their replies and retweets. In Twitter, the replies could also contain replies, we took this into consideration during the collection. The retweets were up to 200 per tweets due to the limitation of the Twitter API. 

\textbf{User Network:} After we acquired all the social engagements of the news stories and news releases, we built a user crawler to get the information of users and construct a network of the involved users. The involved users include users who tweet, reply, retweet and the users who are mentioned in these tweets. The side information of the users consists of the user profiles and user timelines which are recent two hundred tweets. The network could be built through the user followings list and user followers list.

\begin{table}[t]
    \small
    \centering
    \caption{Description of the features including in the dataset.}
    \label{tab:des_dataset}
    \vskip -1em
    \begin{tabularx}{0.98\columnwidth}{p{0.32\columnwidth}X}
    \toprule
    Category & Features  \\
    \midrule
    \multirow{2}{*}{News Contents} & 
    URL, Title, Key words, Text,   \\
    & Images, Tags, Authors, Date\\
    \midrule
    \multirow{5}{*}{News Reviews} & Rating of the News, \\ 
     & Ground Truth of Rating Criteria, \\
     & Explanations of the Ground Truth, \\
     & Tags, Category, Title, Summary, \\
     & Descriptions, Images, News Source \\
     \midrule
     Social Engagements & Tweets, Replies, Retweets,\\
     \midrule
     \multirow{2}{*}{User Network}
     & User Profiles, User Timelines,\\
     & User Followings, User Followers\\
    \bottomrule
    \end{tabularx}
    \vspace{-3mm}
\end{table}

\setlength{\tabcolsep}{15pt}
\begin{table*}[!htb]
    \small
    \centering
    \caption{The statistics of the datasets.}
    \label{tab:statistics}
    \vskip -1em
    \begin{tabularx}{0.98\textwidth}{|p{0.25\textwidth}|X|X|X|X|X|X|}
    \hline
     &  \multicolumn{3}{|c|}{HealthStory} & \multicolumn{3}{|c|}{HealthRelease}\\
     \hline
                                        
    & Total       & Real        & Fake        & Total & Real & Fake \\
    \hline                                
    News                                        
    & 1,690       & 1,218       & 472         & 606   & 315  & 291  \\
    \hline
    Tweets count                   
    & 384,073      & 289,731      & 94,342       &47,338       &25,091      &22,247 \\
    Average tweets per news         
    & 227.26      & 237.87      & 199.88      &78.12       &79.65      &76.45      \\
    Average tweets per user per news
    & 1.21        & 1.20        & 1.23        &1.14     &1.13      &1.14      \\
    \hline
    Replies count                               
    & 27,601      & 20,644       & 6,957      &1,575       &685      &890      \\
    Average replies per news                    
    & 16.33       & 16.95       & 14.74       &2.60     &2.17      &3.06      \\
    Average replies per tweets                  
    & 0.072       & 0.071       & 0,074       &0.033    &0.027     &0.040      \\
    \hline
    Retweets count                              
    & 120,709      & 92,758        & 27,951   &16,959      &9,594      &7,365      \\
    Average retweets per news                   
    & 71.43       & 76.16        & 59.22      & 27.99 & 30.46 & 25.31      \\
    Average retweets per tweets                 
    & 0.314       & 0.320        & 0.296      &0.358       &0.382      &0.331      \\
    \hline
    Unique users count 
    & 241,368     &195,425       &70,407      &30,362      &18,474     &15,551 \\
    \hline
    \end{tabularx}
    \vspace{-2mm}
\end{table*}

\section{Dataset Description}
The collected dataset repository \textit{FakeHealth} consists of two datasets, i.e., \textit{HealthStory} and \textit{HealthRelease} corresponding to news stories and news releases. News stories are reported by news media such as Reuters Health, while news releases are from various institutes including universities, research centers and companies. Due to the difference in sources, we group news story and news release into separate datasets in FakeHealth. The information contained in HealthStory and HealthRelease datasets can be categorized into four categories, i.e., news contents, news reviews, social engagements and user networks. Each category has multiple features. We list the details in Table~\ref{tab:des_dataset}.

On the HealthNewsReview.org, both news stories and news releases are evaluated by experts on 10 criteria. Among them, 8 criteria (C1-C8) are common for both datasets. The remaining two are specially designed for HealthStory (S9-S10) and HealthRelease (R9-R10). These criteria assess the health news in diverse aspects such as the overclaiming, missing of information, reliability of sources and conflict of interests. The  contents of the criteria are listed in Table \ref{tab:criteria} in Appendix. Furthermore, each criterion is accompanied with the label and detailed explanations. 
The overall rating score is in proportion to the number of criteria satisfied by the news. The rating score ranges from 0 to 5. Following the strategy in \cite{shu2018fakenewsnet}, we treat news pieces whose scores lower than 3 as fake news. The statistics of the collected datasets are shown in Table \ref{tab:statistics}.

\begin{figure}[t]
\small
\centering
    \begin{subfigure}[t]{0.48\columnwidth}
        \centering
        \includegraphics[width=\linewidth]{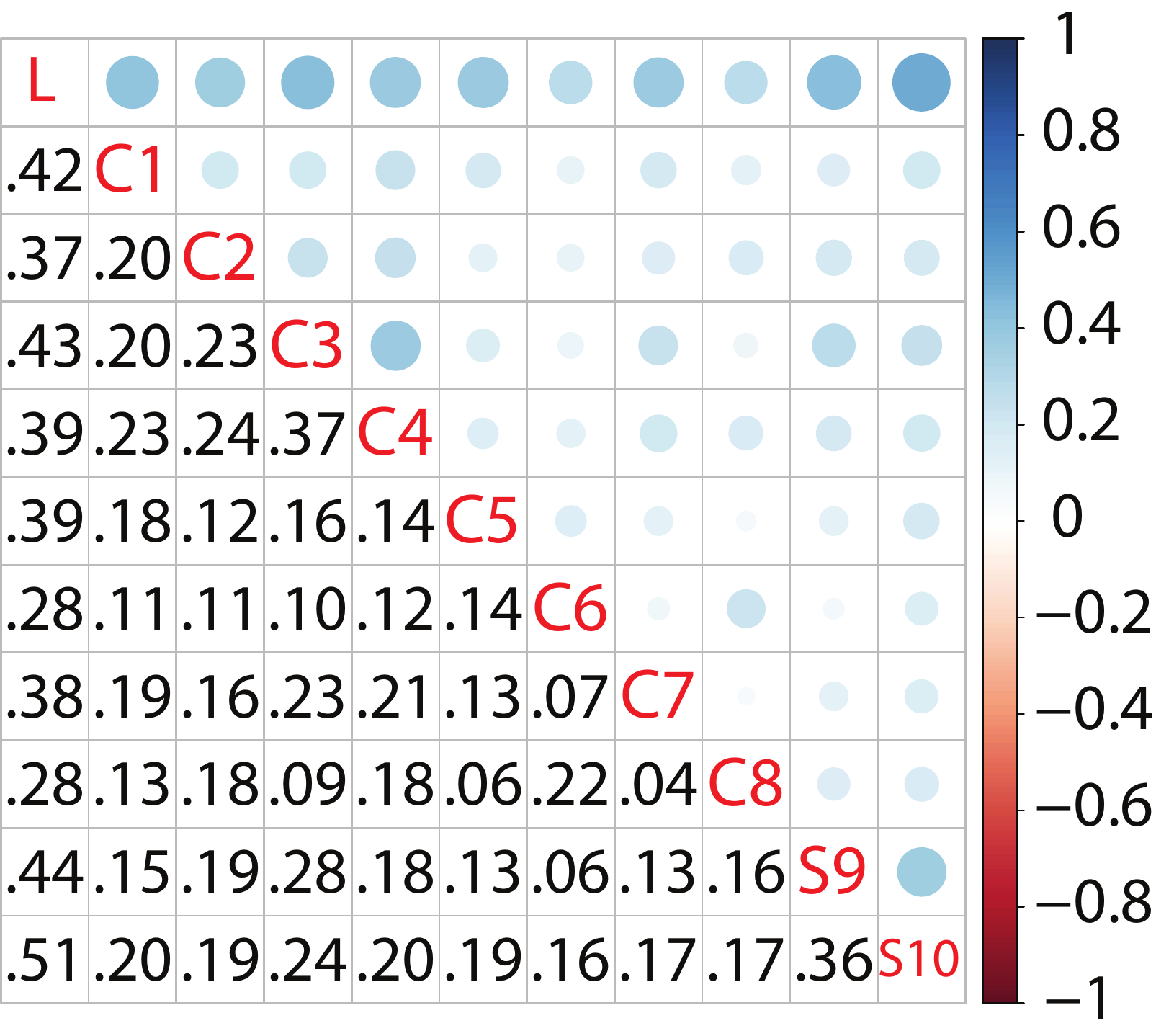}
        \vspace{-5mm}
        \caption{HealthStory}
    \end{subfigure}
    \begin{subfigure}[t]{0.48\columnwidth}
        \centering
        \includegraphics[width=\linewidth]{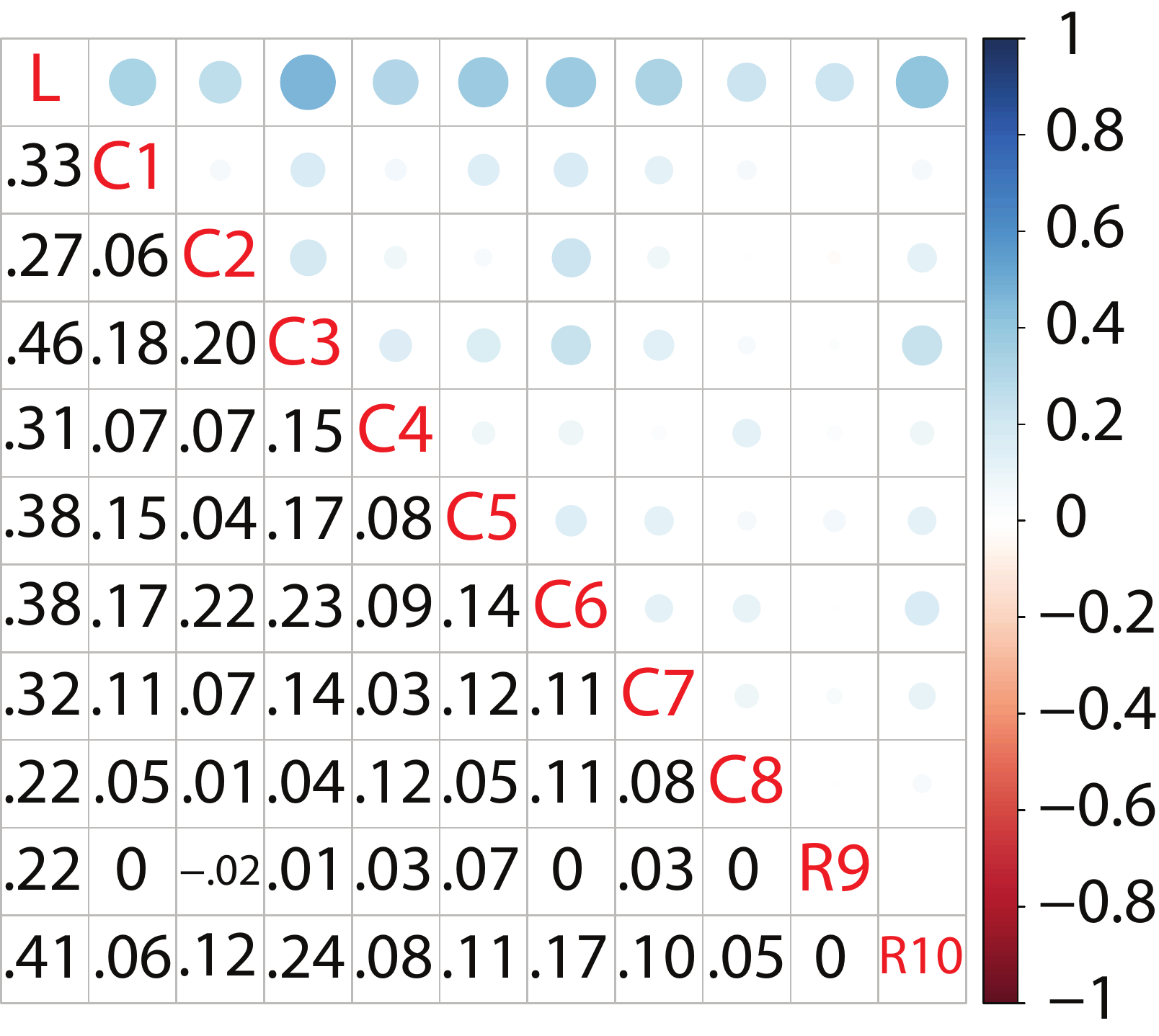}
        \vspace{-5mm}
        \caption{HealthRelease}
    \end{subfigure}
    \vspace{-2mm}
    \caption{The correlation matrices showing the correlation coefficients between the criteria (C1-C10, S9-S10 and R9-R10) and the labels (L) on HealthStory and HealthRelease. }
    \label{fig:cor}
    \vspace{-3mm}
\end{figure}
\setlength{\tabcolsep}{5pt}
\begin{table*}[t]
    \small
    \centering
    \caption{The AUC scores of the criteria prediction by logistic regression with unigram and POS features.}
    \label{tab:cretiera_class}
    \vskip -1em
    \begin{tabularx}{0.98\textwidth}{|p{0.12\textwidth}|X|X|X|X|X|X|X|X|X|X|X|X|}
        \hline
        Dataset       &  C1 & C2 & C3 & C4 & C5 & C6 & C7& C8 & S9 & S10 & R9 &R10\\
        \hline
        HealthStory   &  0.641 & 0.644 & 0.651 & 0.667 & 0.576 & 0.764 & 0.607 & 0.846 & 0.694 & 0.770 & - & -\\
        \hline
        HealthRelease & 0.682 & 0.694 & 0.707 & 0.768 & 0.524 & 0.543 & 0.624 & 0.861& - & - & 0.720 & 0.570\\
        \hline
    \end{tabularx}
    \vspace{-3mm}
\end{table*}

\section{Exploratory Analysis}

In this section, we conduct exploratory analysis to demonstrate the quality of the HealthStory and HealthRelease. 
Besides, we compare fake and real health news in multiple dimensions, which suggests the potential features and challenges of their applications in fake health news detection.  We also provide baselines for future researches.

\subsection{News Reviews}
One of our major contributions is providing the reliable ground truth labels, judgements of  evaluation criteria and corresponding explanations. Here, we perform the analysis to demonstrate the importance and quality of them.

In our datasets, the label of a single criterion is determined by experts, and the ground truth labels to indicate the news pieces as real or fake are based on the number of criteria that they satisfy. This strategy relies on the assumption that \textit{criteria represent independent aspects to assess health news}. To verify this assumption, we investigate the correlation coefficients between the criteria and labels on both HealthStory and HealthRelease. The results are displayed in Figure \ref{fig:cor}. We observe a minimal positive or even no correlation between the criteria. We also find that the ten criteria contribute equally to the label. These two observations confirm the assumption. The independence between the criteria also implies that health news could be fake for many reasons.

In order to show the judgements of the criteria are valid to be learned by algorithm, we applied a linguistic-style based model to predict whether the health news satisfies the ten criteria. The model is a logistic regression classifier with unigram and part of speech (POS) features. We randomly sample 100 pieces of positive samples and negative samples for each criterion as test set. Table \ref{tab:cretiera_class} displays the results. From the table, we observe that with logistic regression an simple features of unigram and POS, we can have relatively good AUC for most of the criteria (AUC of random guess should be 0.5). This indicates the possibility of predicting why the news is fake based on the news content. The review criteria can potentially facilitate novel research tasks such as explainable fake news detection and multi-aspect news rating.

\subsection{News Topics}
A fair fake news dataset should have similar topic distributions across the fake category and real category. If the dataset doesn't comply with it, the classifier trained on the datset is likely to be a topic classifier instead of fake news detector. Thus, we adopt two approaches to verify that both HealthStory and HealthRelease have similar topic distribution across fake and real category.

\begin{figure}[t]
\centering
    \begin{subfigure}[t]{0.48\columnwidth}
        \centering
        \includegraphics[width=\linewidth]{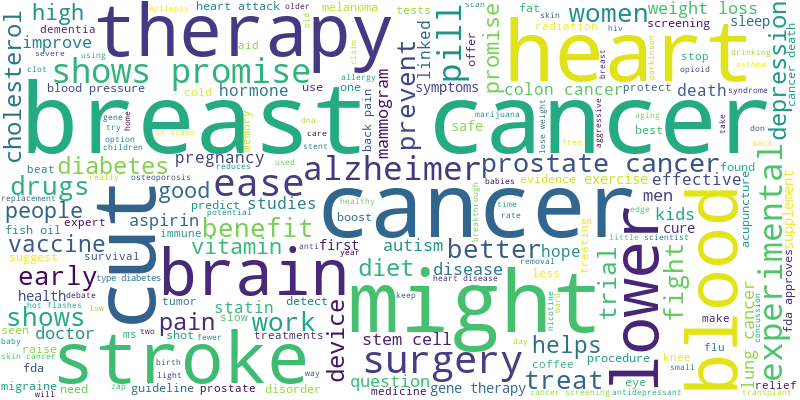} 
        \label{fig:wc_a}
        \vspace{-4mm}
        \caption{Real HealthStory}
    \end{subfigure}
    \begin{subfigure}[t]{0.48\columnwidth}
        \centering
        \includegraphics[width=\linewidth]{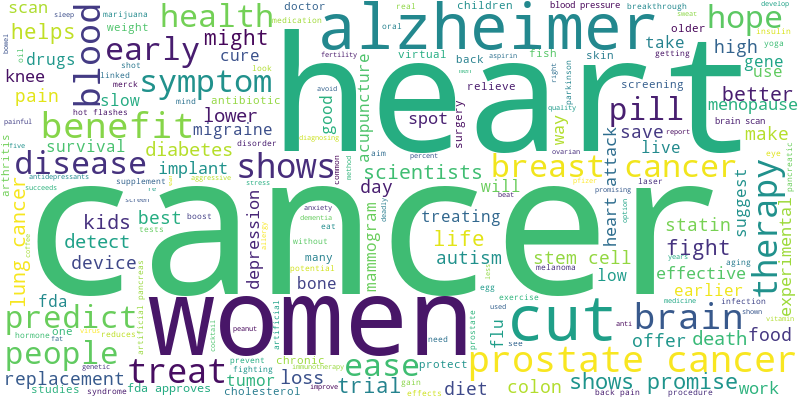} 
        \label{fig:wc_b}
        \vspace{-4mm}
        \caption{Fake HealthStory}
    \end{subfigure}
    \begin{subfigure}[t]{0.48\columnwidth}
        \centering
        \includegraphics[width=\linewidth]{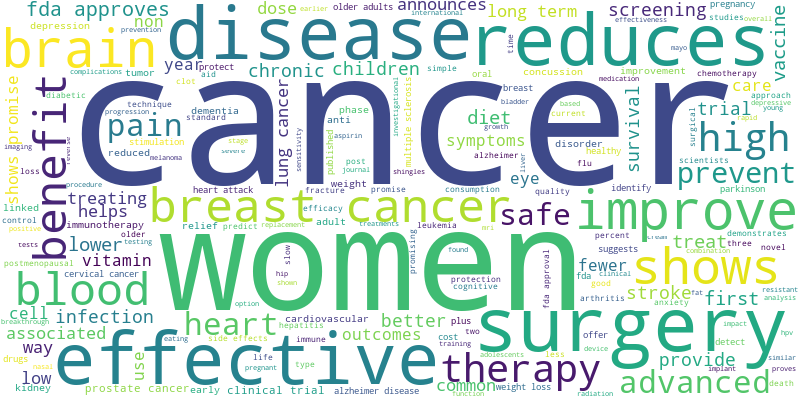} 
        \label{fig:wc_c}
        \vspace{-4mm}
        \caption{Real HealthRelease}
    \end{subfigure}
    \begin{subfigure}[t]{0.48\columnwidth}
        \centering
        \includegraphics[width=\linewidth]{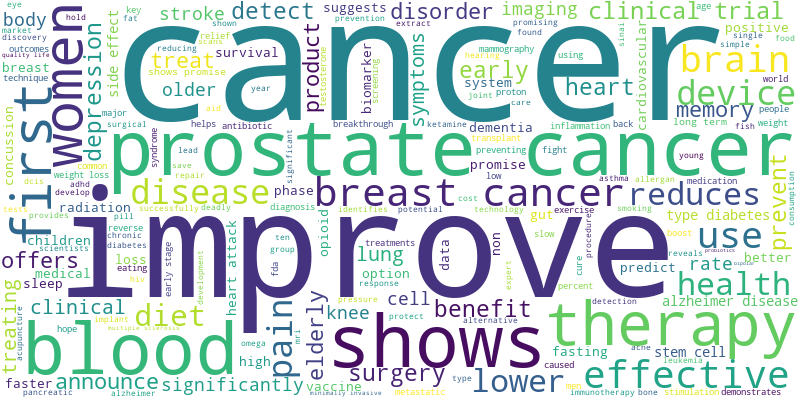} 
        \label{fig:wc_d}
        \vspace{-4mm}
        \caption{Fake HealthRelease}
    \end{subfigure}
    \vspace{-2mm}
\caption{The word cloud of news headline for real and fake news on HealthStory and HealthRelease.}
\label{fig:wordcloud}
\vspace{-3mm}
\end{figure}

We first visualize the most frequent words of news headlines with word cloud to compare the topics between the fake news and real news, which is shown in Figure \ref{fig:wordcloud}.  We observe that real news and fake news of HealthStory share most of the frequent words such as cancer, heart and alzheimer. We have similar observation in HealthRelease, i.e., both real and fake news of HealthRelease mainly focus on cancer, surgery and therapy. This implies that the topics of true and fake news are consistent for both datasets.

\begin{figure}[t]
\centering
\begin{subfigure}{0.48\columnwidth}
    \centering
    \includegraphics[width=\linewidth]{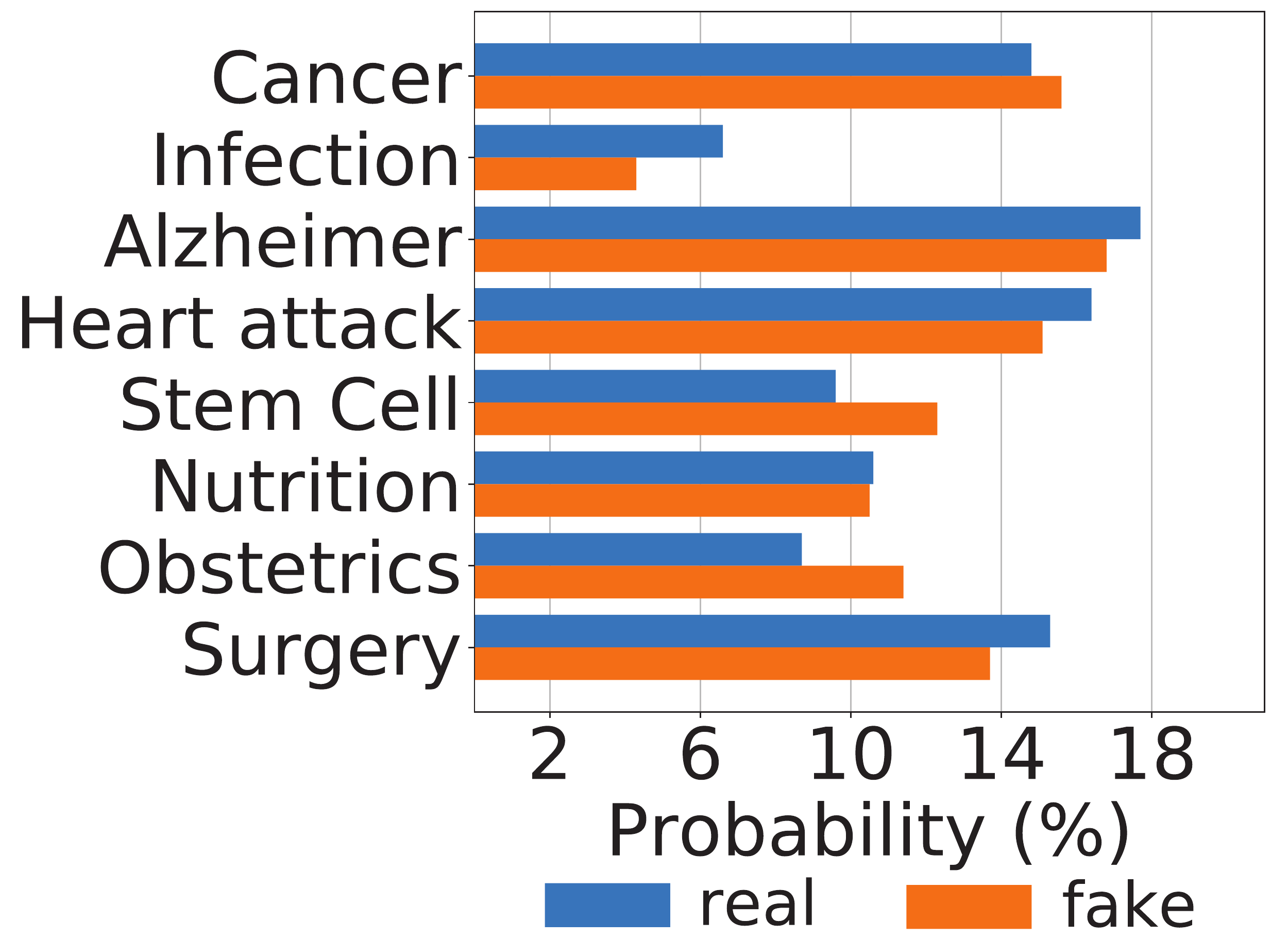} 
    \vspace{-5mm}
    \caption{HealthStory}
    \label{fig:story_LDA}
\end{subfigure}
\begin{subfigure}{0.48\columnwidth}
    \centering
    \includegraphics[width=\linewidth]{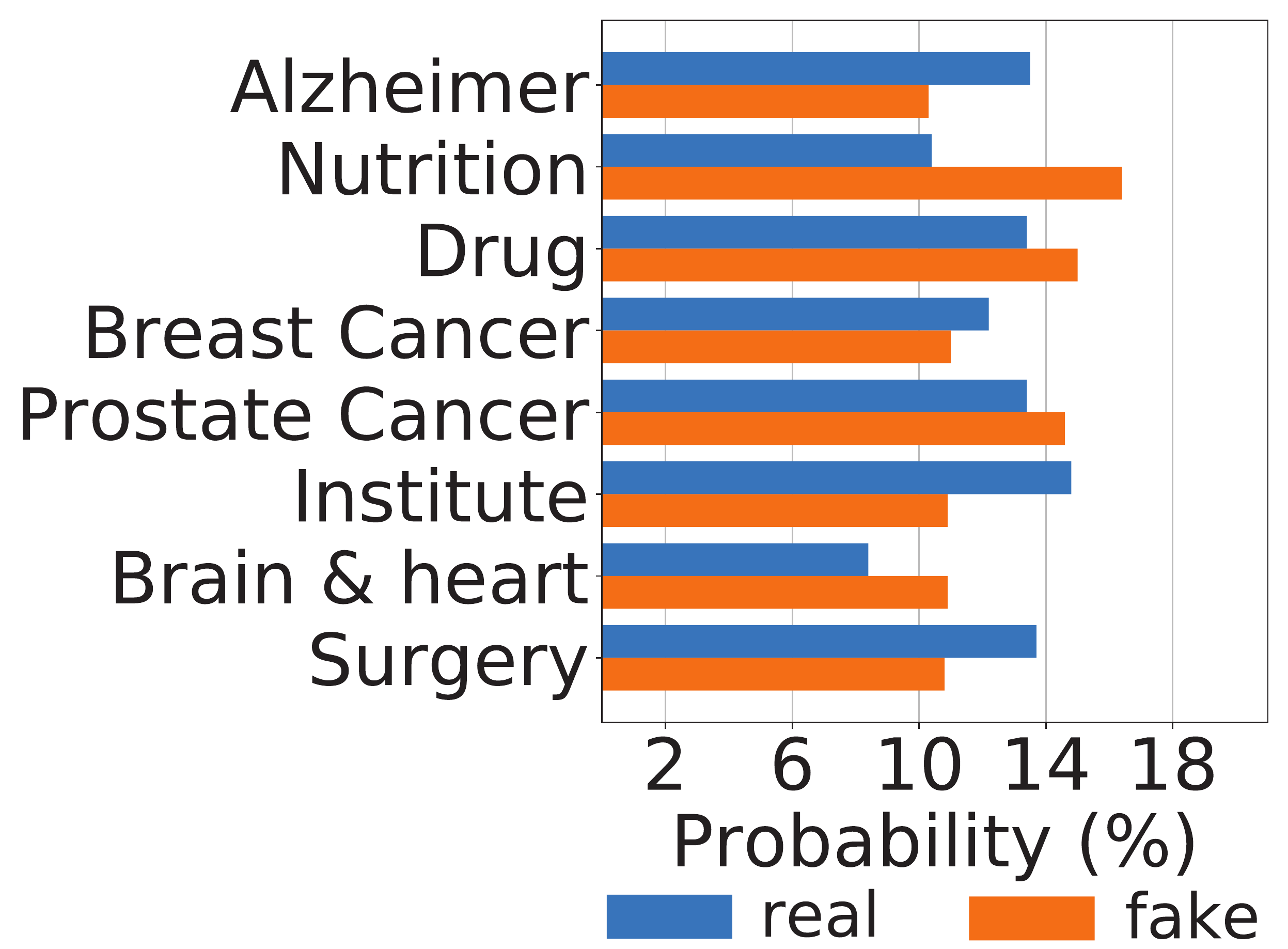} 
    \vspace{-5mm}
    \label{fig:news_LDA}
    \caption{ HealthRelease}
\end{subfigure}
\vspace{-2mm}
\caption{The distributions of the topics for real and fake news on HealthStory and HelathRelease.}
\vspace{-3mm}
\label{fig:LDA}
\end{figure}

To quantitatively analyze the topics across real and fake news pieces, we explore the topic distributions with latent Dirichlet allocation (LDA) \cite{blei2003latent}. LDA is a topic modeling algorithm, which assumes a document is generated from a mixture of topics. Topics represented by a distribution of words and the topic distributions of the documents are learned after the training of LDA. 
We set the number of topics in LDA as eight, with reference to HealthReviewNews.org. To make the topics found by LDA more interpretable, we assign a word or phase to represent each topic.
The topic distribution comparisons of real and fake articles are shown in Figure \ref{fig:LDA}. The quantitative results of LDA is in line with the visualization of word cloud. There are marginal difference between real and fake news in topic distributions.
The accordance of real and fake news topic distributions ensure the fake news detection models trained on our datasets are not topic classifiers. 

We further explore the topics that differ relatively noticeably in LDA analysis. It appears that a few topics are more likely to involve fake news, which results in this difference. For instance, despite the wide spread of news about stem cell therapies, most of the them are unproven by U.S. Food and Drug Administration except one for blood production disorder. This fact accounts for the unbalanced number of the fake and real news around stem cells. This evidence suggests that news tags and key words may provide valuable information for fake news detection in health domain.

\begin{figure}[t]
    \centering
    \includegraphics[width=0.9\columnwidth]{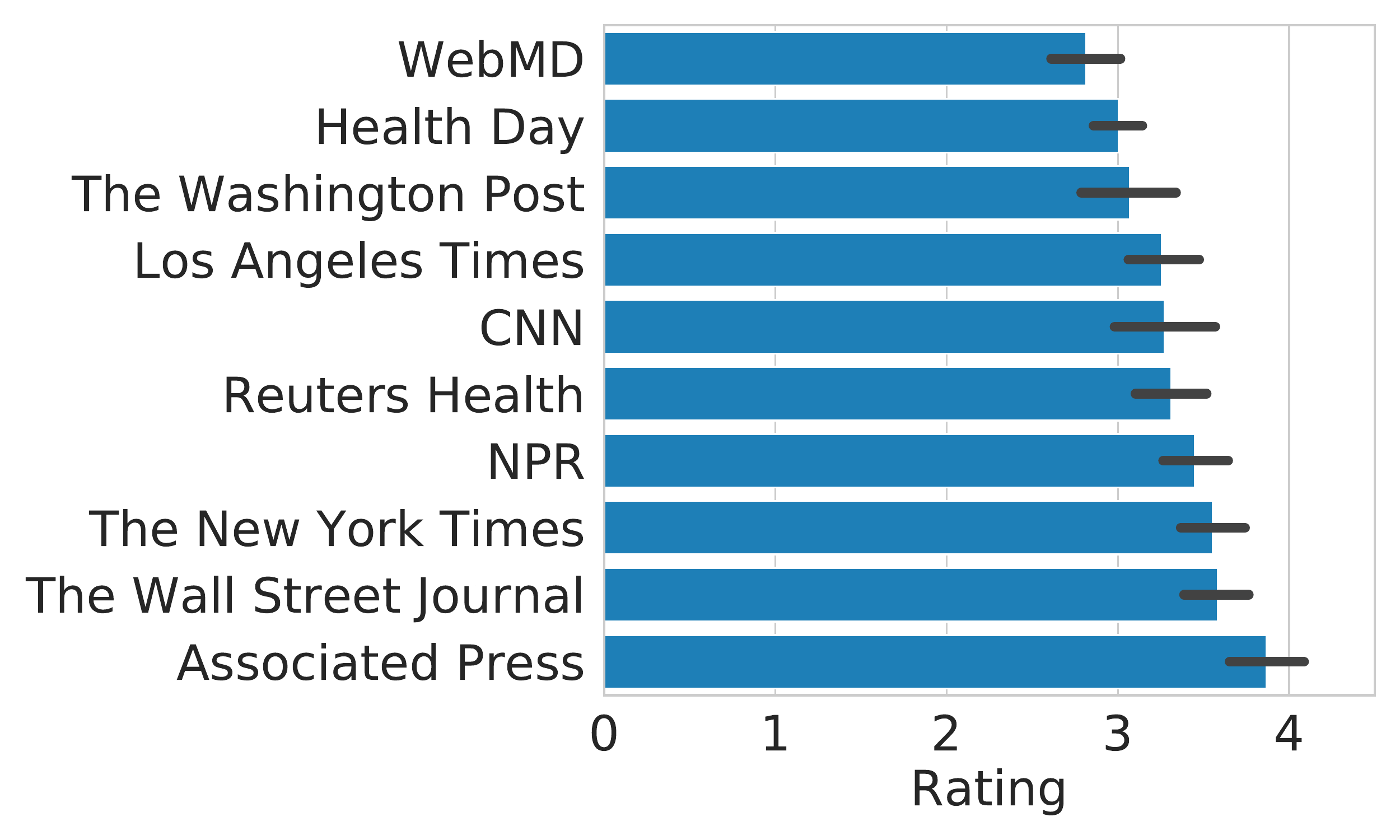}
    \vspace{-3mm}
    \caption{The average ratings of news from different sources in HealthStory. The error bars denote 95\% confidence interval of the means.}
    \label{fig:source}
    \vspace{-3mm}
\end{figure}

\subsection{News Sources}
The news sources could serve as auxiliary information for fake news detection \cite{zhou2018fake,shu2017fake}. On some occasions, the fake news detection method is even simplified as website source recognition. For instance,  \citet{silverman2016analysis} straightforwardly treats news from fake news websites or hyperpartisan websites that present themselves as publishing real news as fake news.
To confirm whether the similar trend exists among health news sources, we evaluate the news sources credibility with the average ratings of the publishers. Because in HealthRelease, the publishers are too diverse to have meaningful comparison, we only analyze the sources in HealthStory dataset.  We choose news agencies that have more than 50 pieces news reviewed in our dataset. From Figure \ref{fig:source}, we find that the average ratings differ a lot among publishers. The 95\% confidence interval of the average ratings don't overlap for many news publishers. For instance, Associated Press's average rating is as high as 3.87, while WebMD's average rating is 2.82. The rating of Associated Press's news story is significantly larger than WebMD ($p<0.001$, t-test). Our observation indicates that news sources could be a useful feature, which is consistent with existing work.

\subsection{Social Context}

In this section, we analyze the characteristics of FakeHealth in the dimensions of tweets, retweets, replies, user credibility and temporal engagements patterns. 

\begin{figure}[t]
\centering
    \begin{subfigure}[t]{0.48\columnwidth}
        \centering
        \includegraphics[width=\linewidth]{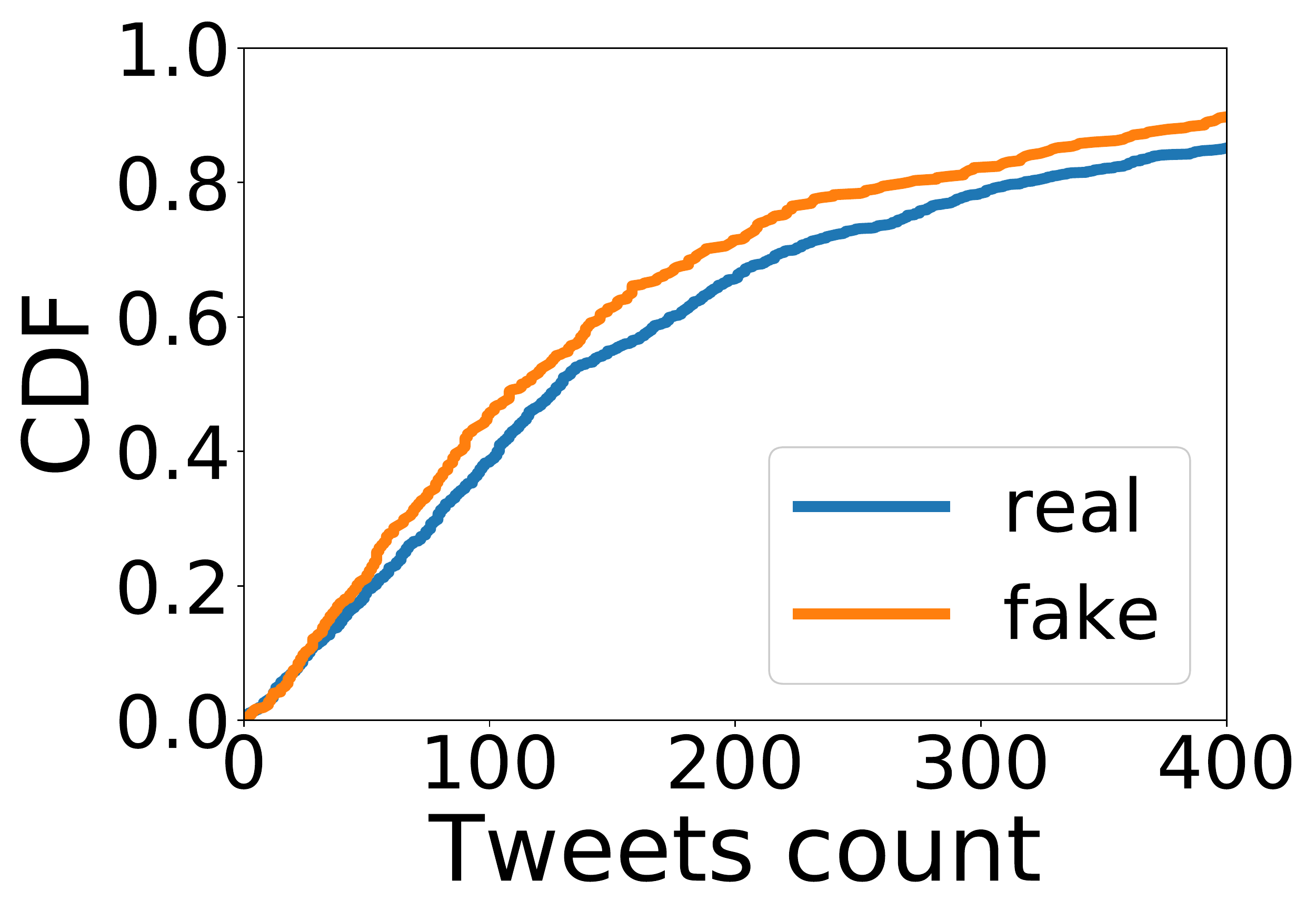} 
        \vspace{-5mm}
        \caption{HealthStory}
        \label{fig:story_tweets_dis}
    \end{subfigure}
    \begin{subfigure}[t]{0.48\columnwidth}
        \centering
        \includegraphics[width=\linewidth]{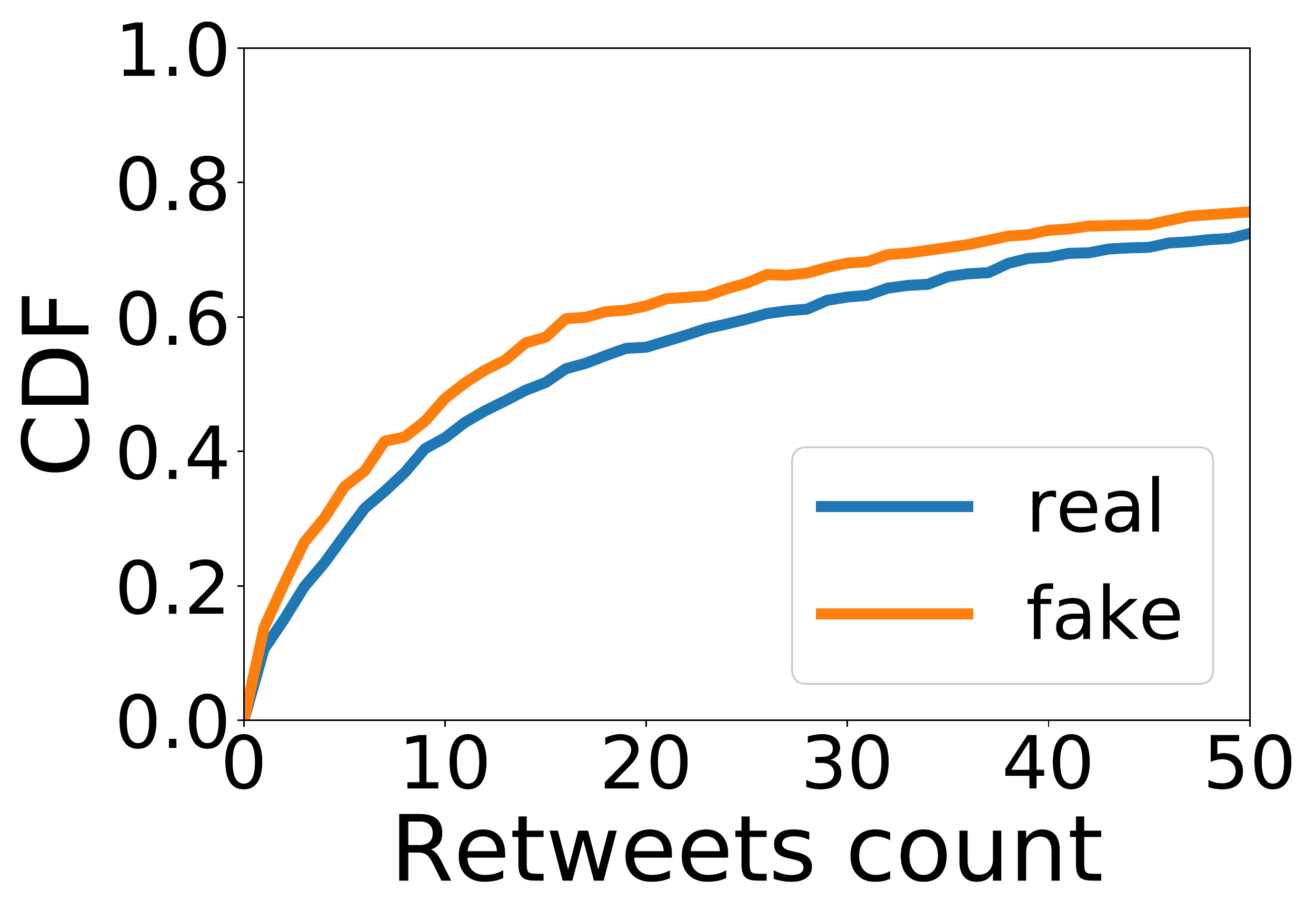} 
       \vspace{-5mm}
        \caption{HealthStory}
        \label{fig:story_retweets_dis}
    \end{subfigure}
    
    \begin{subfigure}[t]{0.48\columnwidth}
        \centering
        \includegraphics[width=\linewidth]{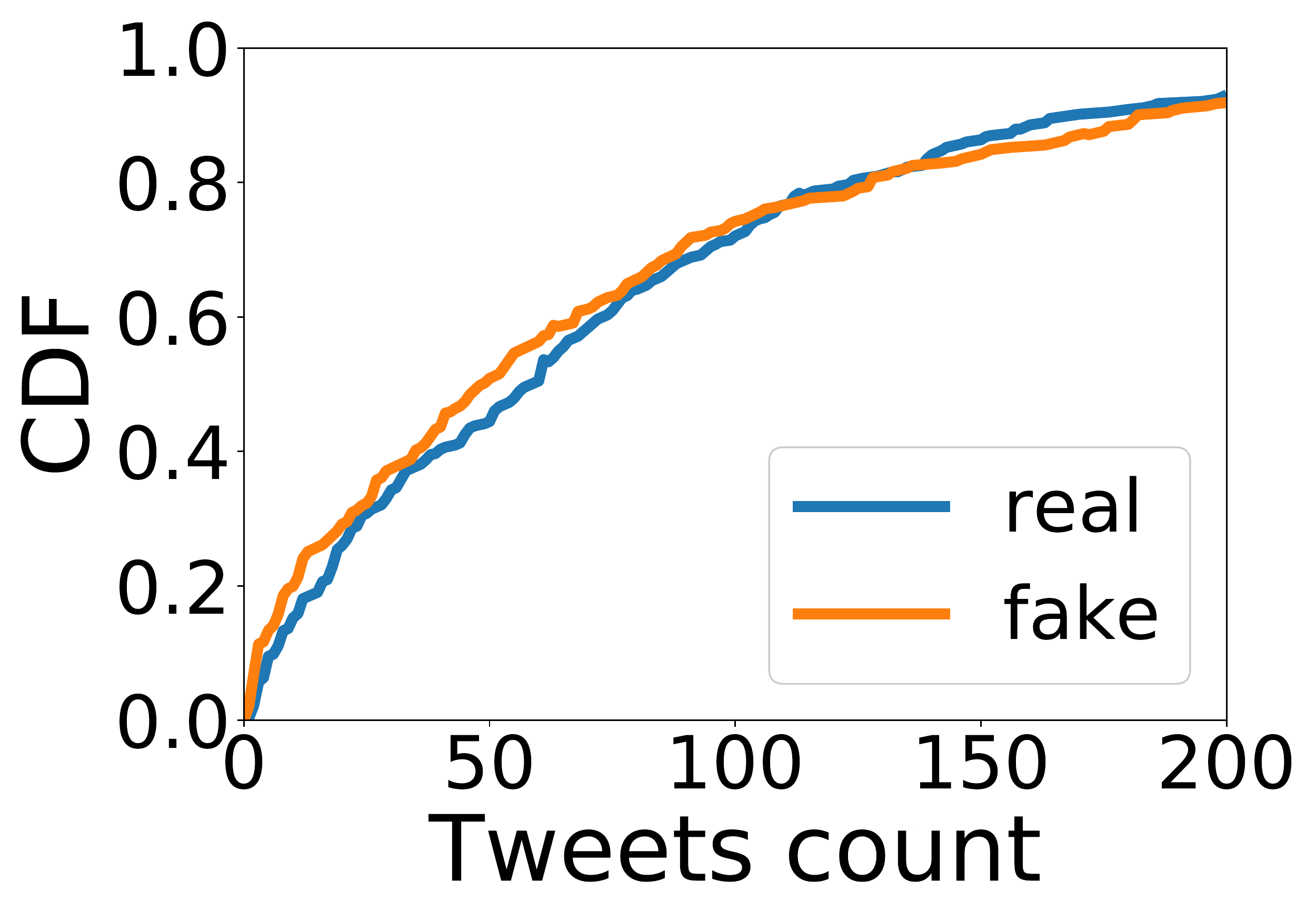}
        \vspace{-5mm}
        \caption{HealthRelease}
        \label{fig:news_tweets_dis}
    \end{subfigure}
    \begin{subfigure}[t]{0.48\columnwidth}
        \centering
        \includegraphics[width=\linewidth]{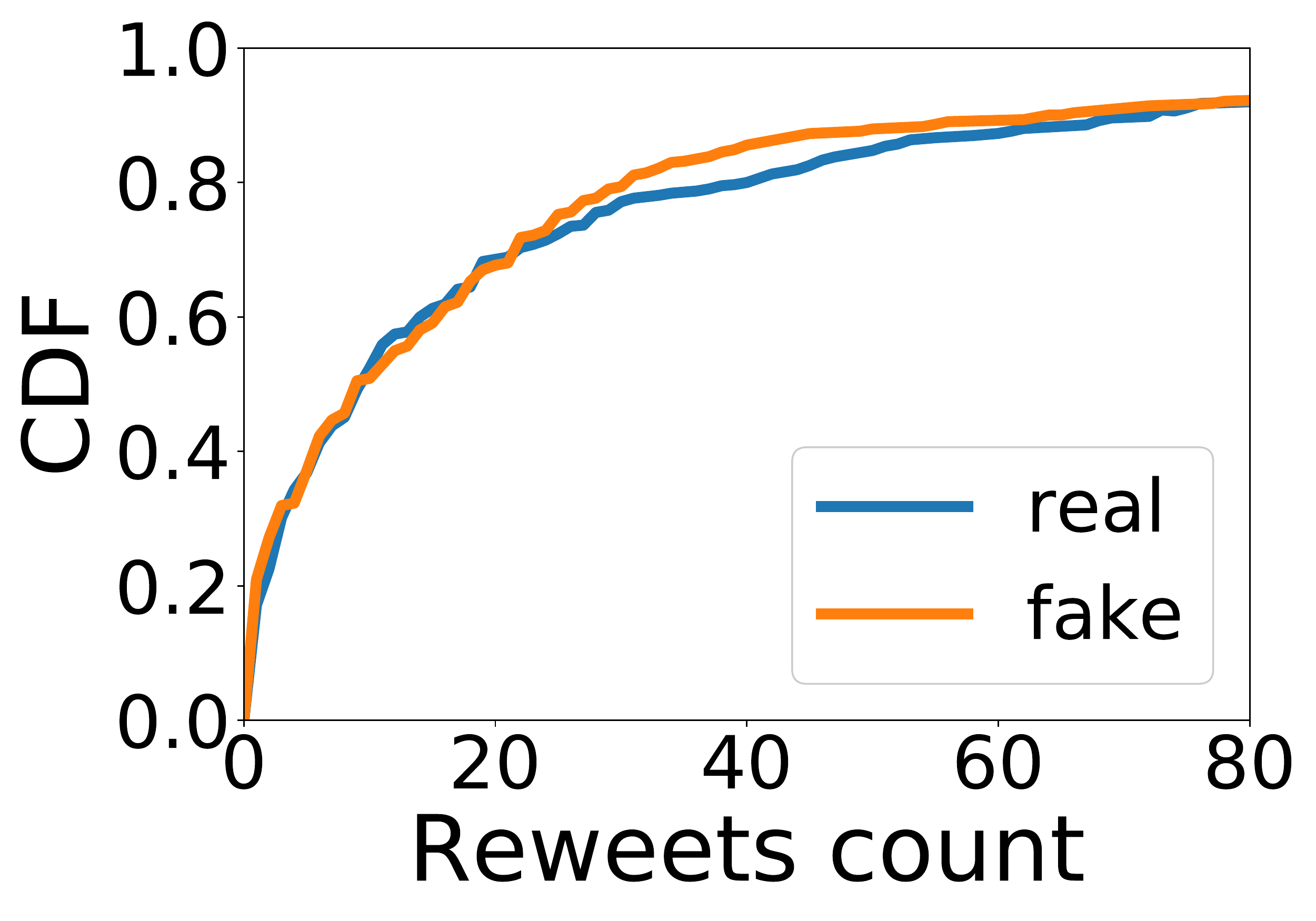}
        \vspace{-5mm}
        \caption{HealthRelease}
        \label{fig:news_retweets_dis}
    \end{subfigure}
    \vspace{-2mm}
\caption{Cumulative distribution functions of real and fake news in HealthStory and HealthRelease.  }
\label{fig:count}
\vspace{-3mm}
\end{figure}

\textbf{Tweets and Retweets:}
The numbers of tweets posting health news and its retweets indicate the social impact of the news piece. Here we compare the distributions of tweets count and retweets count between real and fake news to find whether the impact of real health news is higher than the fake one. 
The results displayed in Figure \ref{fig:count} (\subref{fig:story_tweets_dis}) and (\subref{fig:story_retweets_dis}) show differences between fake and real news in HealthStory. Both the tweet count and retweet count of real news story are larger than fake news ($p<0.001$, Mann-Whitney U test).
This observation in HealthStory is consistent with some earlier work \cite{friggeri2014rumor,kumar2016disinformation}. \citet{friggeri2014rumor} show true rumors on Facebook will have more reshares than false rumors. \citet{kumar2016disinformation} demonstrate the hoaxes on Wikipedia are viewed less than non-hoaxes.
However,  there is no significant difference in tweet count and retweet count between real news and fake news in HealthRelease. Actually, in HealthStory, the impact of fake news is also very high. These observations reveal the challenge of fake news detection in health domain.

\textbf{Replies:} Users express their view points towards the original news by replies. The replies may oppose or support the news. These conflicting voices could be viewed as crowd source assessments. Thus, user replies are often utilized in the social context-based approaches such as SAF \cite{shu2019fakenewstracker}.
Here we perform sentiment analysis showing that the replies towards real news are more positive. 
We obtain the polarity scores through a state-of-the-art model for social media text sentiment analysis, i.e., VADER~\cite{hutto2014vader}. The polarity scores consist of four parts: compound score, negative score, neutral score and positive score. Compound score is a sentiment metric normalized to between -1 (most negative) and +1 (most positive). The negative, neutral and positive scores represent the fractions of the text that fall in each category. The average polarity scores of the replies are listed in Table \ref{tab:sentiment_scores}. As expected, the compound scores of news in HealthStory are significantly higher than fake one ($p<0.001$, Mann-Whitney U test). In addition, the negative scores and neutral scores of real stories are significantly less ($p<0.001$, Mann-Whitney U test). In HealthRelease, we observe the similar trend. However, the differences are not statistically significant. This probably due to the  relatively limited number of replies in HealthRelease. To summarize, the observation in HealthStory implies that replies might be helpful to identify fake health news.



\setlength{\tabcolsep}{2pt}
\begin{table}[t]
\small
\centering
\begin{tabularx}{0.98\columnwidth}{p{0.30\columnwidth}p{0.18\columnwidth}XXX}
\toprule
Dataset                      & Compound & Negative & Neutral & Positive \\
\midrule                         
Real HealthStory   & 0.142  & 0.064  & 0.789  & 0.148   \\
Fake HealthStory   & 0.079  & 0.069  & 0.810  & 0.121   \\
\midrule
Real HealthRelease & 0.172   & 0.045   & 0.802  & 0.152   \\
Fake HealthRelease & 0.161   & 0.046   & 0.803  & 0.151   \\ 
\bottomrule
\end{tabularx}
\caption{Comparison of average polarity scores of replies to tweets posing real and fake news.}
\label{tab:sentiment_scores}
\vspace{-3mm}
\end{table}

\begin{figure}[t]
    \begin{subfigure}[t]{0.48\columnwidth}
        \centering
        \includegraphics[width=\linewidth]{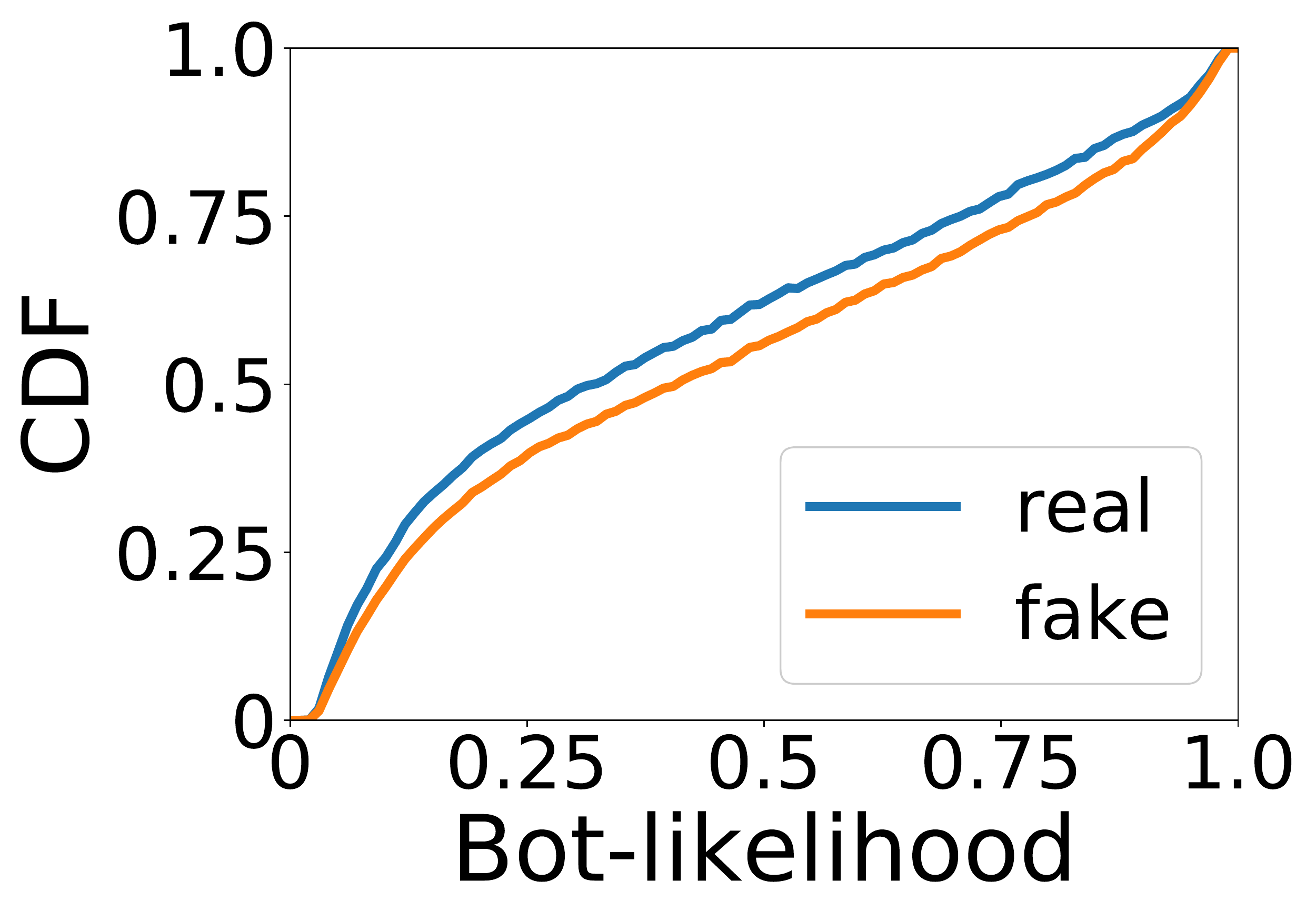}
        \vspace{-5mm}
        \caption{HealthStory}
    \end{subfigure}
    \begin{subfigure}[t]{0.48\columnwidth}
        \centering
        \includegraphics[width=\linewidth]{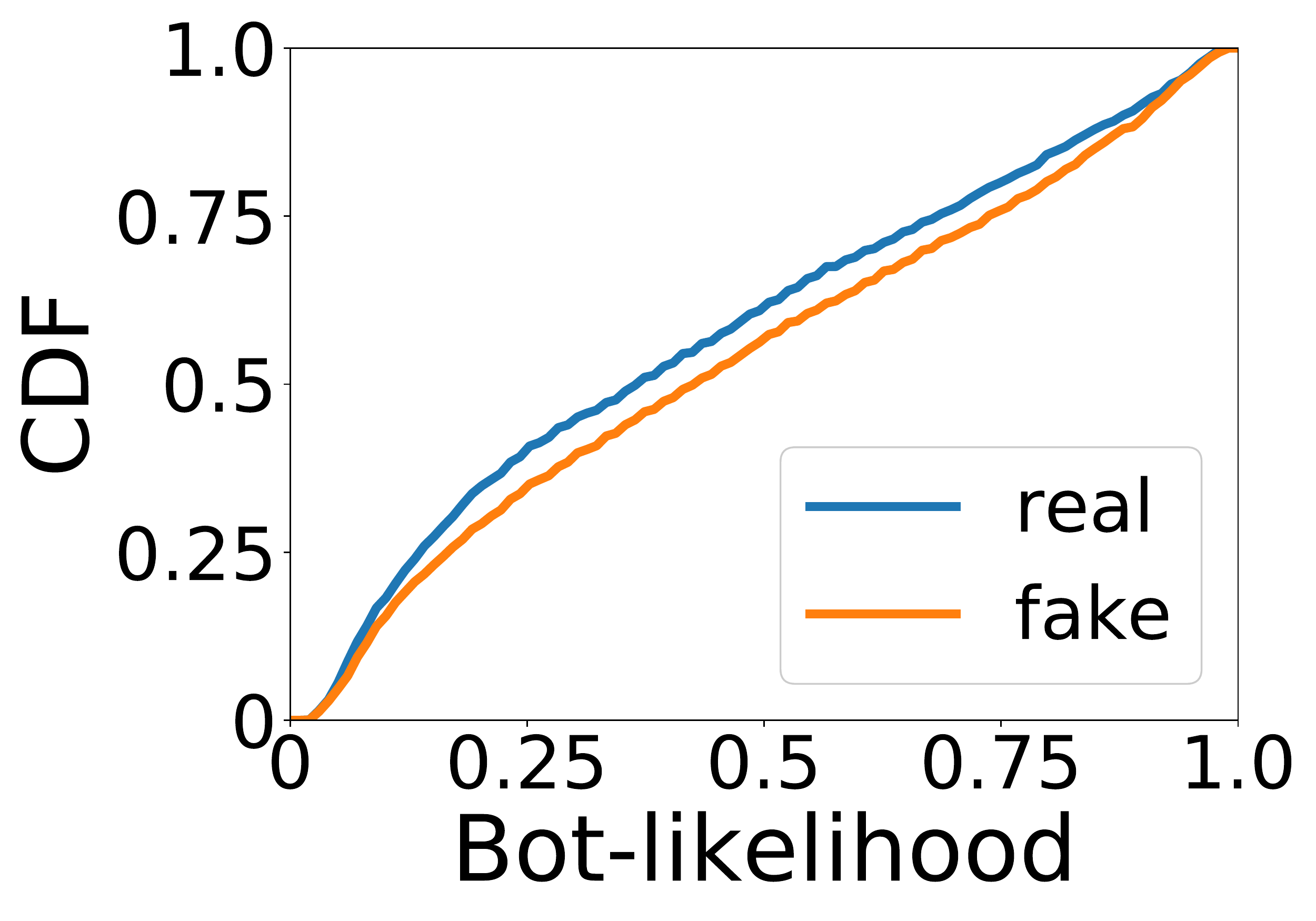}
        \vspace{-5mm}
        \caption{HealthRelease}
    \end{subfigure}
    \vspace{-2mm}
    \caption{Cumulative distribution functions showing the different bot-likelihood distributions for users involved in real and fake health news.}
    \label{fig:bot_likehood}
    \vspace{-3mm}
\end{figure}

\textbf{User Credibility:}
Users on the social media have different credibility levels based on their behaviors. According to \citet{abbasi2013measuring}, users with low credibility are responsible for part of misinformation. There have been several studies which consider user credibility in fake news detection algorithms \cite{gupta2012evaluating,shu2019beyond}. 
Here, we investigate whether the users involved with health news show equal credibility by estimating their bot-likelihood scores. Additionally, we contrast real health news with fake health news in related user bot-likelihood scores to confirm that the propagation of fake health news is more likely involved with users in low credibility level.   
We randomly sample 2000 users who are only involved with real news and 2000 users related with fake news in both HealthStory and HealthRelease datasets. The bot-likelihoods of the users are obtained by querying the user names through the BotoMeter API\footnote{https://botometer.iuni.iu.edu/}. 
BotoMeter~\cite{davis2016botornot} is a state of the art bots detection algorithm based on the features of network, user, friend, temporal, content and sentiment. Figure \ref{fig:bot_likehood} shows the distributions of two groups of user bot-likelihood scores with cumulative distribution function. Similar with the observations in \cite{vosoughi2018spread,shu2018fakenewsnet}, in both HealthStory and HealthRelease, the users who propagate fake news are slightly more likely to be bots. Furthermore, it is obvious that users involved with the health news are not in the same credibility level. It is more reasonable to consider the user credibility when we identify fake news based on users behaviors.



\begin{figure}[t]
    \centering
    \begin{subfigure}[t]{0.48\columnwidth}
        \centering
        \includegraphics[width=\linewidth]{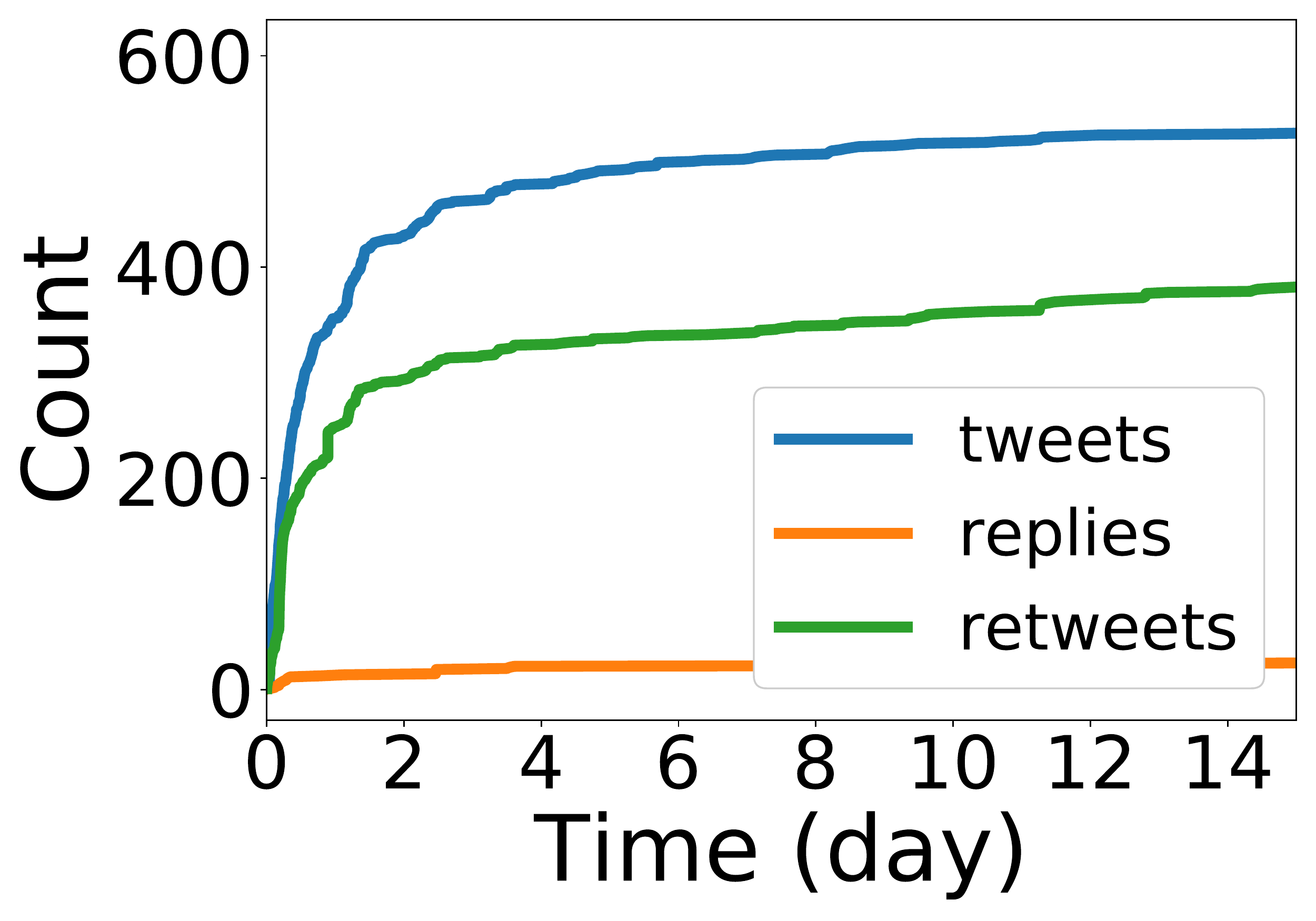}
        \vspace{-5mm}
        \caption{Real health news}
        \label{fig:real_temppral_pattern}
    \end{subfigure}
    \begin{subfigure}[t]{0.48\columnwidth}
        \centering
        \includegraphics[width=\linewidth]{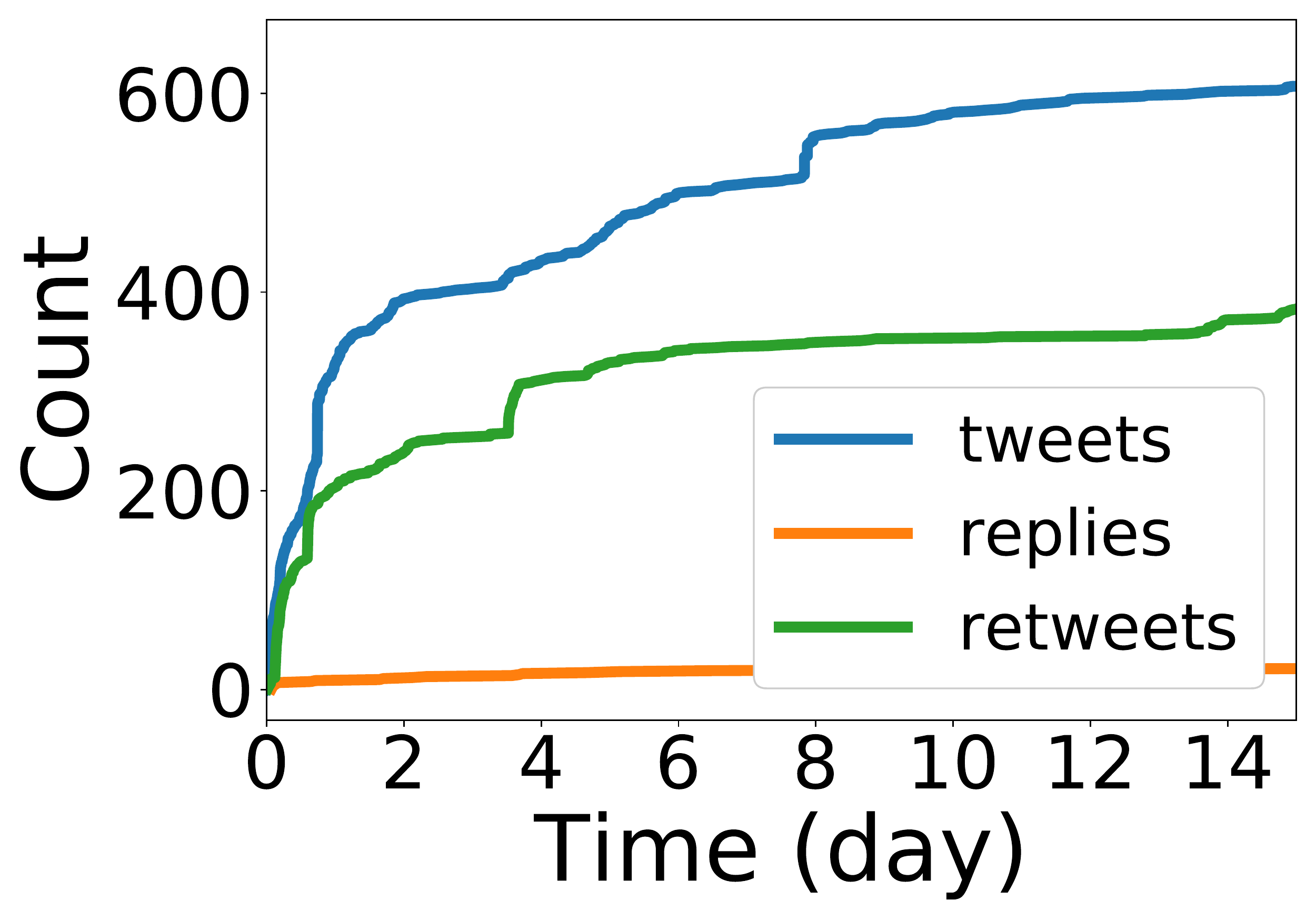}
        \vspace{-5mm}
        \caption{Fake health news}
        \label{fig:fake_temppral_pattern}
    \end{subfigure}
    \vspace{-2mm}
    \caption{Temporal patterns of social engagements of fake health news and real health news.}
    \label{fig:Temporal pattern}
    \vspace{-3mm}
\end{figure}

\begin{figure}[t]
\centering
    \begin{subfigure}[t]{0.48\columnwidth}
        \centering
        \includegraphics[width=\linewidth]{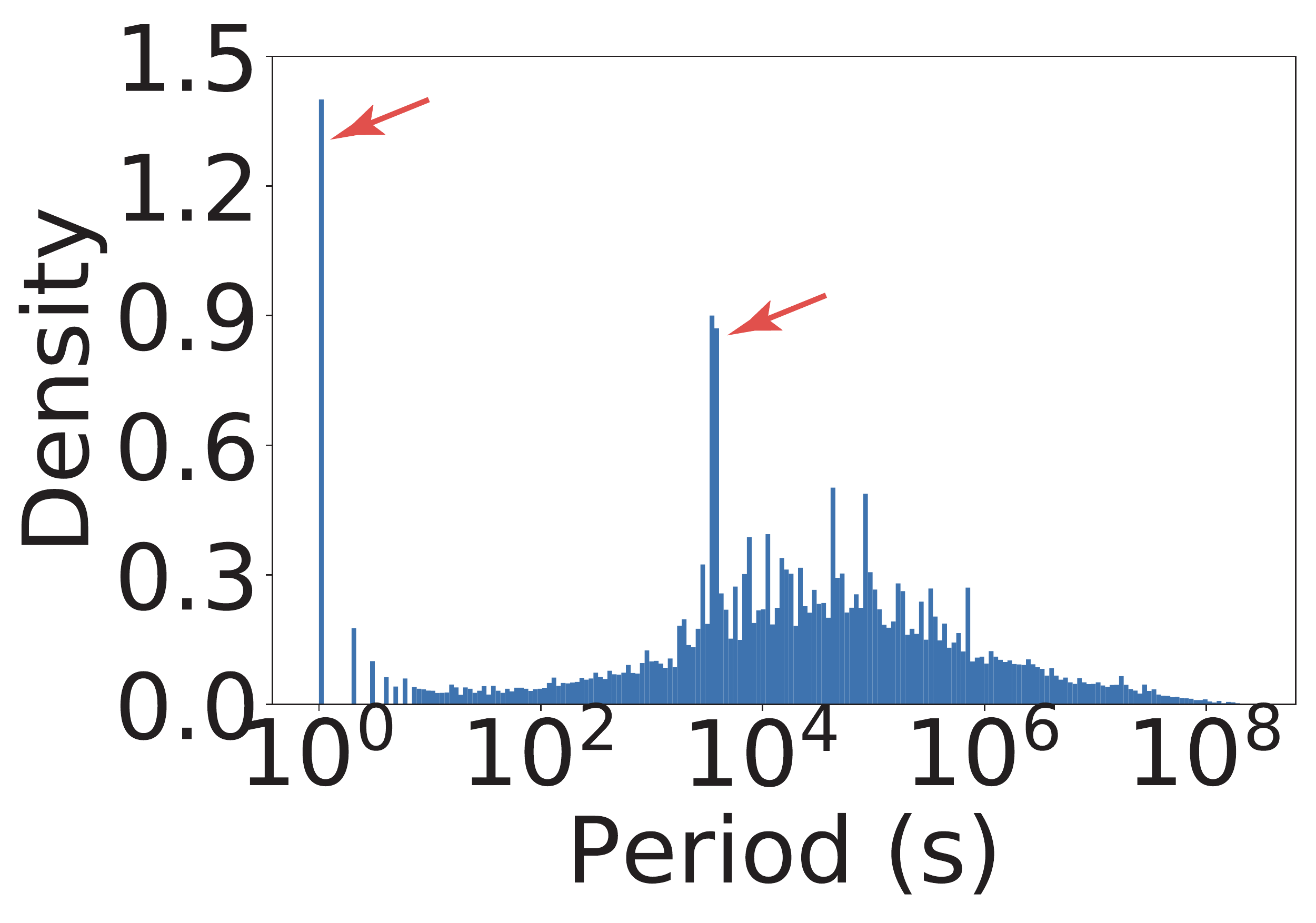}
        \vspace{-5mm}
        \caption{HealthStory}
    \end{subfigure}
    \begin{subfigure}[t]{0.48\columnwidth}
        \centering
        \includegraphics[width=\linewidth]{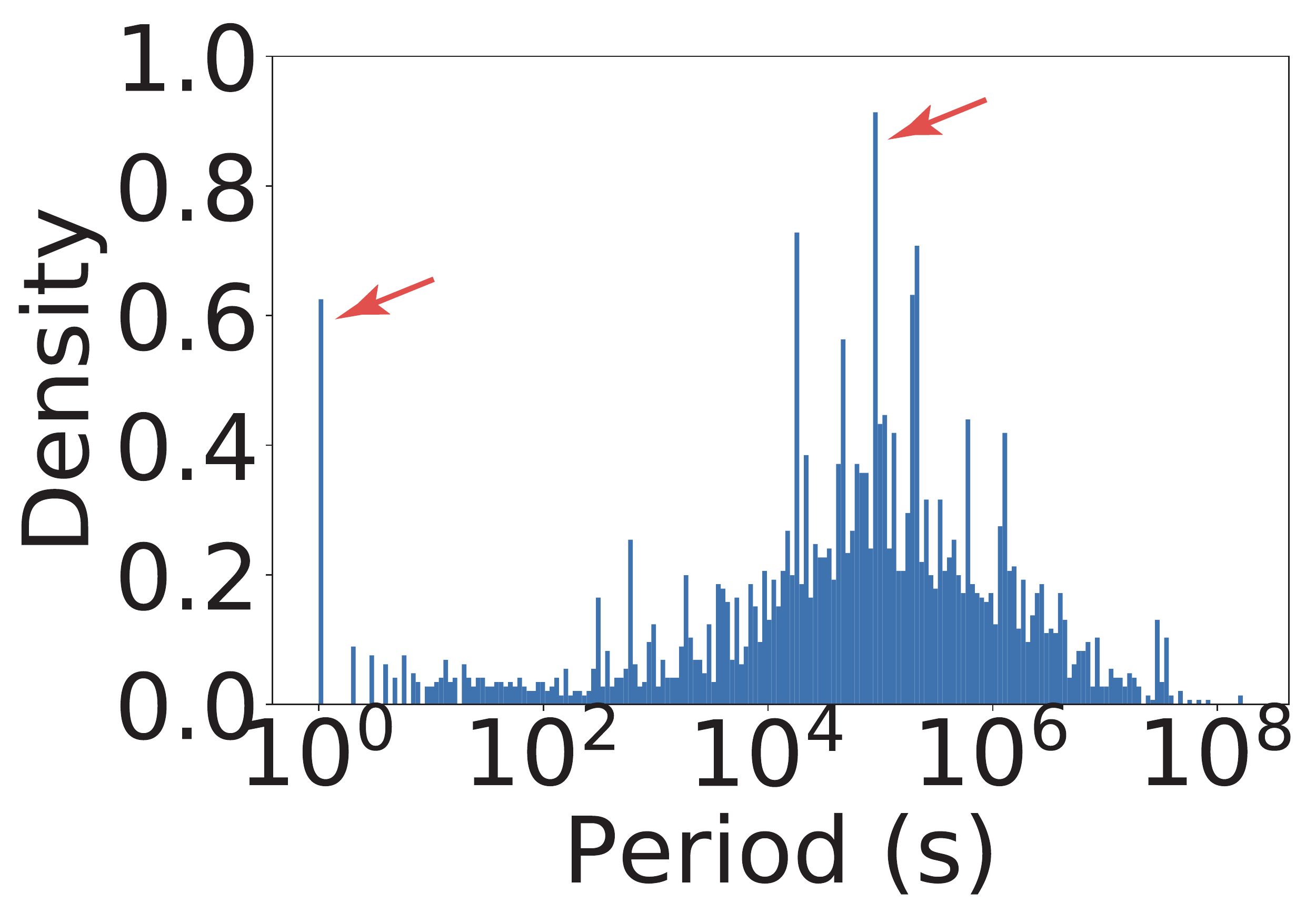}
        \vspace{-5mm}
        \caption{HealthRelease}
    \end{subfigure}
    \vspace{-2mm}
    \caption{The distributions of tweeting periodicity in HealthStory and HealthRelease. Peaks are marked by arrows.}
    \label{fig:period}
    \vspace{-3mm}
\end{figure}

\begin{figure}[t]
\centering
    \begin{subfigure}[t]{\columnwidth}
        \centering
        \includegraphics[width=\linewidth]{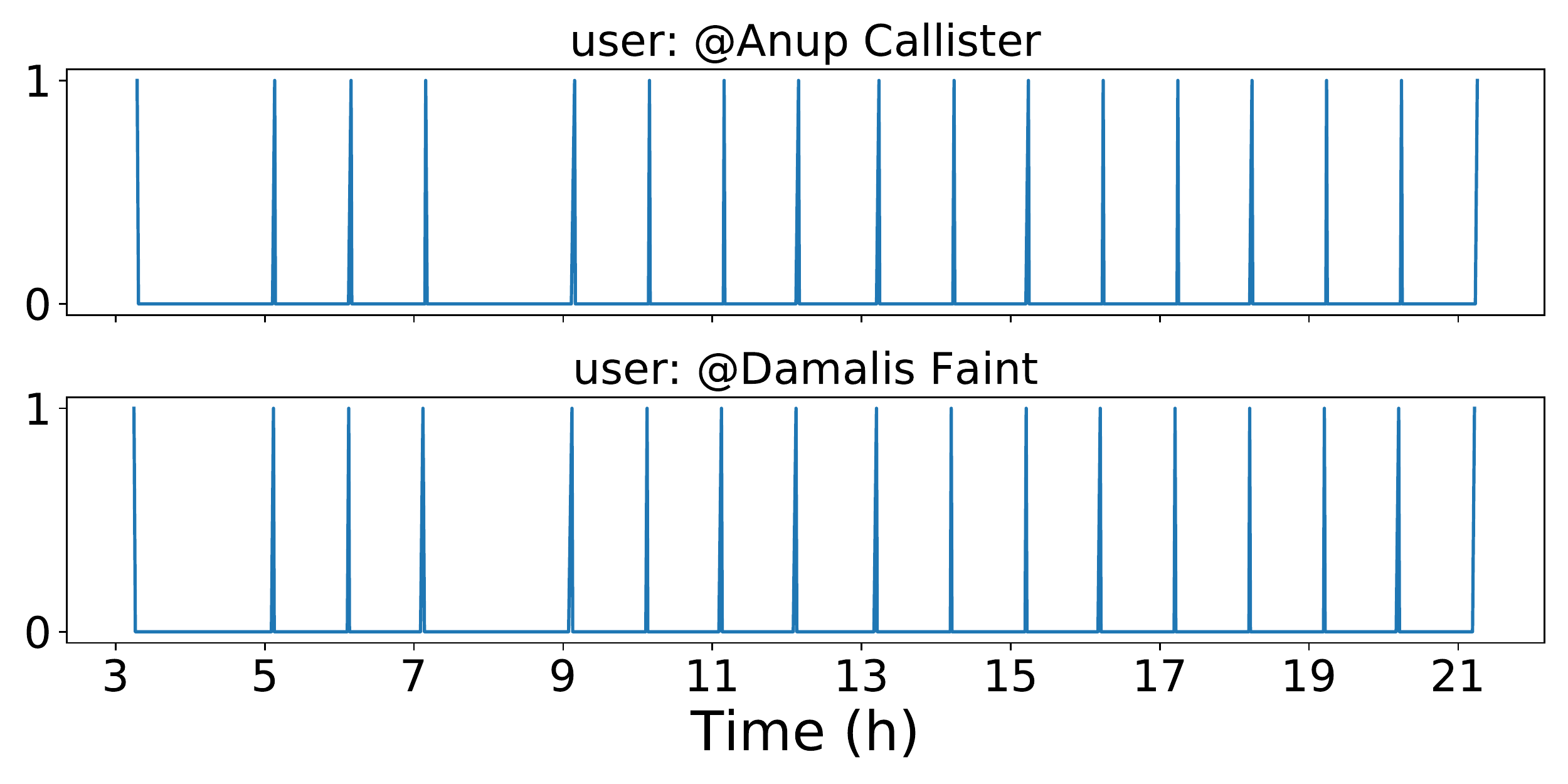}
        \vspace{-6mm}
        \caption{}
        \label{figure:pattern_cor}
    \end{subfigure}
 
    \begin{subfigure}[t]{\columnwidth}
        \centering
        \includegraphics[width=\linewidth]{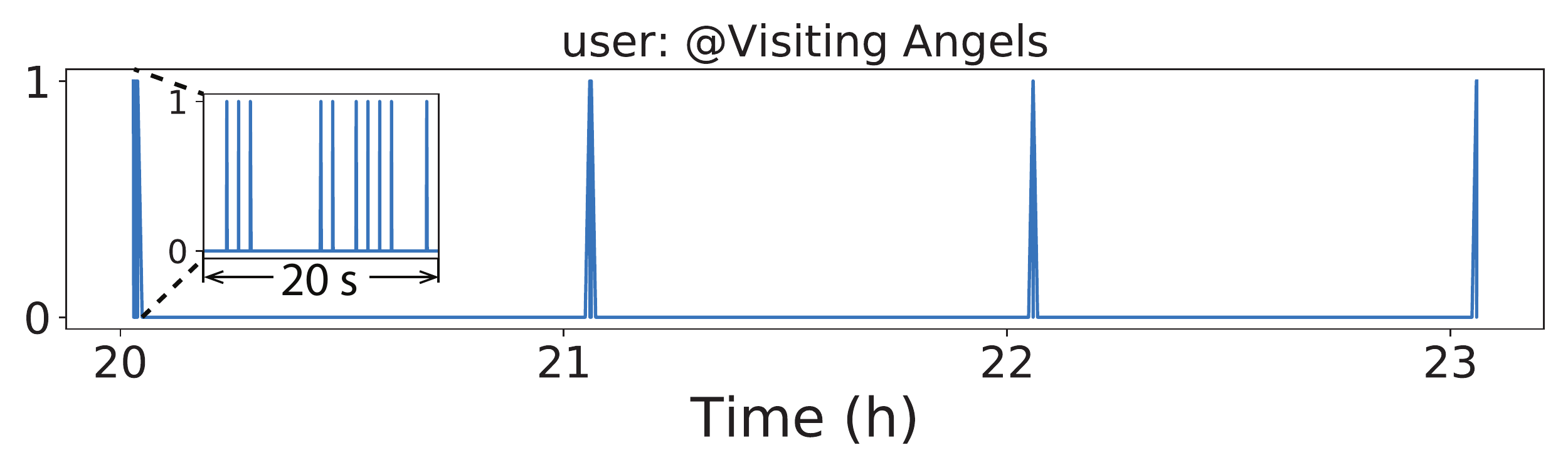}
        \vspace{-6mm}
        \caption{}
        \label{figure:pattern_burst}
    \end{subfigure}
    \begin{subfigure}[t]{\columnwidth}
        \centering
        \includegraphics[width=\linewidth]{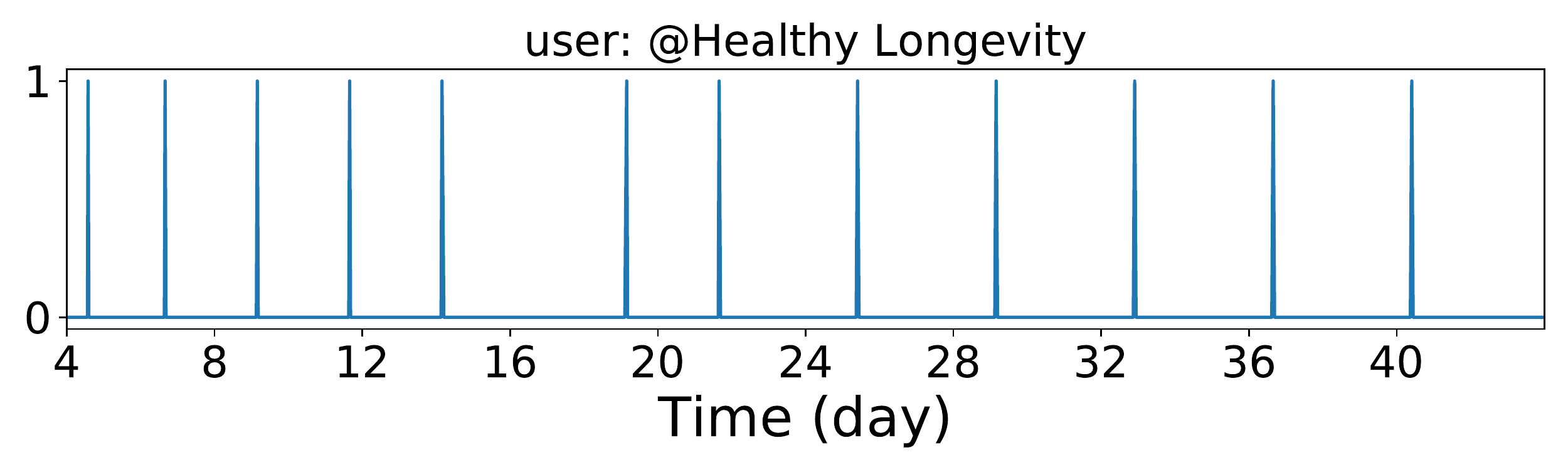}
        \vspace{-6mm}
        \caption{}
        \label{figure:pattern_long}
    \end{subfigure}
\vspace{-4mm}
\caption{Three periodic patterns of bots activities in health news propagation. x-axis shows the time after the news published, y-axis shows the number of tweets. (a) Periodically tweeting from users showing perfect temporal correlation. (b) Periodically tweeting a burst of tweets. (c) Tweeting fake health news with a long period.}
\label{fig:pattern}%
\vspace{-3mm}
\end{figure}

\textbf{Temporal Engagement Pattern:}
Temporal information of social engagements has been explored in a number of fake news detection methods \cite{ma2016detecting,shu2019fakenewstracker}. 
To check whether temporal patterns of health news social engagements could be utilized to detect fake health news, we examined two news pieces in our datasets. From Figure \ref{fig:Temporal pattern}, we observe a sudden increase of tweets and retweets for the fake news. On the contrary, real news social engagements increase steadily. Thus, the spike in the temporal engagements pattern can be an indicator of fake news.

Considering our observation that users posting fake health news have higher bot-likelihood scores, the burst of tweeting and retweeting may be related to the bots. To investigate how bots involve with the propagation of health news, we present the bots activities in our datasets in two dimensions. First, we conduct a periodicity detection on both HealthStory and HealthRelease as suggested in \cite{chavoshi2017temporal}.  The periods are evaluated by the delay of the successive same tweets from the same user. Figure \ref{fig:period} reveals that two peaks of the period distributions exist in both HealthStory and  HealthRelease. One of the peak represents very short delay around 1 second, which is also in agreement with the pattern shown in Figure \ref{fig:Temporal pattern} (\subref{fig:fake_temppral_pattern}).  Another peak is around several days.
Second, we also find three special periodic patterns of bots activities, which are:
\begin{enumerate}
    \item Several users periodically tweets the same news simultaneously. An example is shown in Figure \ref{fig:pattern} (\subref{figure:pattern_cor}), the activities of the two accounts present perfect temporal correlation, which indicates they are controlled by one host.
    \item Some users periodically post a burst of tweets. As shown in Figure \ref{fig:pattern} (\subref{figure:pattern_burst}), the period of the burst is around 1 hour. The burst lasts round 20 seconds.
    \item In addition to the short period, there are bots that post the news with a long periodicity to have continuous impact. In Figure \ref{fig:pattern} (\subref{figure:pattern_long}), user @Healthy Longevity repeatedly tweets the same fake news more than a month.
\end{enumerate}
These patterns confirm that bots are affecting the social trend of health news with various strategies. And the dissemination of health news could last for several month, while the propagation of general news mostly accomplishes in 4 days \cite{glenski2018propagation}. These patterns and special characteristic may provide insights for the temporal pattern related detection models. 


\subsection{Performance of Fake Health News Detection }

In this subsection, we conduct fake health news detection on the two datasets. We only include very simple baselines because the purpose is not to achieve high performance, but to (\textit{i}) verify the quality of our datasets for fake news detection; (\textit{ii}) provide reference results for future evaluations on the datasets; and (\textit{iii}) show that there are much space to improve the performance by developing state-of-the-art methods and incorporating various information provided by the datasets.

 We adopt three different types of baselines including: (i) linguistic-based, (ii) content-based and (iii) social context-based detection methods.
Unigram is a lexicon-level features for linguistic-based methods \cite{ott2011finding}. Here, we trained a logistic regression classifier with unigram. We also added news source (NS) and tags to the linguistic-based model to verify their potential improvements to detection. To give more references, SVM and random forest are implemented with unigram as baselines as well. Two content-based models, CNN and bidirectional GRU (Bi-GRU), were employed for fake health news detection \cite{kim2014convolutional,chung2014empirical}. SAF corresponds to social context-based method \cite{shu2019fakenewstracker}. It uses the sequence of replies and user embeddings for fake news detection. LSTM is applied to deal with the temporal pattern of these sequential features \cite{gers1999learning}. In SAF,  the social context features learned from a LSTM encoder are combined with the social context features to make the final prediction .

For both HealthStory and HelathRelease, we randomly selected 70\% data as training set and 30\% as test set. Five-fold cross validation is conducted to tune the hyperparameters. The results of the baselines are listed in Table \ref{tab:fake news detection performance}. Overall, the three types of baselines have relatively good performance compared with random guess. This demonstrates that our datasets are valuable for fake news detection. Our results of unigram+NS, unigram +Tags and SAF indicate that utilizing more features in our datasets is helpful. But performance of these simple baselines is still limited and shows a great potential of improvements. Thus, more explorations of novel fake health news detection are required.


In summary, the performance of the baselines validates the quality of labels and collected features in our datasets. It also suggests the demand of further investigation about the features and the potential of developing novel algorithms.

\begin{table}[t]
\small
\centering
    \caption{Fake news detection performance.}
\label{tab:fake news detection performance}
    \vskip -1em
    \begin{tabularx}{0.98\columnwidth}{|p{0.22\columnwidth}|p{0.25\columnwidth}|X|X|X|}
    \hline
    Dataset & Model & Accuracy & F1 & AUC  \\
    \hline
    \hline
    \multirow{9}{*}{HealthStory}
    & Random guess & 0.513 & 0.525 & 0.474 \\
    \cline{2-5}
    & Unigram &0.665  &0.687  & 0.719 \\
    \cline{2-5}
    & Unigram+NS & 0.738 & 0.733 & 0.798 \\
    \cline{2-5}
    & Unigram+Tags & 0.676 & 0.694 & 0.728 \\
    \cline{2-5}
    & SVM & 0.640 & 0.664 & 0.700 \\
    \cline{2-5}
    & Random Forest & 0.720 & 0.735 & 0.790\\
    \cline{2-5}
    & CNN &0.742 &0.730 & 0.814\\
    \cline{2-5}
    & Bi-GRU &0.691  &0.716 &0.772 \\
    \cline{2-5}
    & SAF &0.760  & 0.756 & 0.763 \\
    \hline
    \hline
    \multirow{9}{*}{HealthRelease}
    & Random guess & 0.512 & 0.523 & 0.547\\
    \cline{2-5}
    & Unigram & 0.654 & 0.652 & 0.703  \\
    \cline{2-5}
    & Unigram+NS & 0.665 & 0.667 & 0.712\\
    \cline{2-5}
    & Unigram+Tags & 0.642 & 0.636 & 0.690 \\
    \cline{2-5}
    & SVM &0.657 & 0.674 & 0.706 \\
    \cline{2-5}
    & Random Forest & 0.651 & 0.629 & 0.726\\
    \cline{2-5}
    & CNN & 0.670 & 0.688 & 0.700  \\
    \cline{2-5}
    & Bi-GRU & 0.665 & 0.677 & 0.662  \\
    \cline{2-5}
    & SAF &0.810 & 0.802  &0.809 \\
    \hline
    \end{tabularx}
\vspace{-3mm}
\end{table}

\section{Potential Research Directions}

Our goal is to provide datasets with detailed information for research around health news. With the massive side information obtained for the news pieces, a number of potential applications will benefit from it.

\subsection{Explainable Fake Health News Detection}

In our dataset, every news is provided with ten binary labels to indicate whether it meets the standard health criteria. This unique feature could facilitate the novel research of explainable fake health news detection which is able to recognize the poorly illustrated aspects. Since explanations are given for the identified fake health news, the social media users will understand the reasons behind the classification results. Thus, people are more likely to trust the model and stop sharing the recognized fake health news. Furthermore, it could be a tool to help users think critically towards health news. As a result, the propagation of fake health news could be intervened.
Explainable model also benefits their developers by providing the evidence of overall classification. It enables the researchers to better understand the model and track the problems. For instance, if the training dataset is biased or polluted with adversarial samples, developers are able to recognize the hidden issues based on the analysis on the explanations.    

\subsection{Knowledge-based Fake Health News Detection}

News authenticity assessment by journalism requires to find the related knowledge and evaluate based on evidence instead of language patterns. However, most existing machine learning approaches contrast to this process, which makes them less sensitive to fake news composed of credible sounding sentences. Knowledge graph for health news could potentially address this problem. In our datasets, the news contents and the news reviews post background knowledge, detailed explanations and corrections. With the news contents and reviews, subject-predicate-object triples could be extracted to build knowledge graphs. We could verify or falsify a piece of related health news by comparing with the claimed values and the retrieved values from the graph. It could give explanations based on the retrieved knowledge, which makes it a more promising direction.

\subsection{Credibility-based Fake Health News Detection}
Here, we point out that the credibility of multiple items in our datasets is valuable for fake health news detection. First, some topics are much more likely to be false in health domain. For instance, only one stem-cell based therapy is proven by U.S. Food and Drug Administration until 2019, but it is evident that a large number of unproven therapies are recommended to clinics\footnote{https://www.fda.gov/consumers/consumer-updates/fda-warns-about-stem-cell-therapies}. This is line with the observation in Figure \ref{fig:LDA}, which shows the news about stems cells more likely to be fake. Collected tags and key words can meet the need of adding topic as features for prediction. Second, our analysis presented in \ref{fig:source} shows quality of health news is highly related to the news sources, which implies the credibility of news sources should be useful for spotting fake health news. Third, social bots could be identified based on their tweets, profiles and friends \cite{davis2016botornot}, which suggests that these features contain the necessary information to evaluate the user credibility. And our datasets supply these features of users involved with health news. Thus, algorithms that implicitly consider user credibility by adding user side information could be developed. 

\subsection{Multi-Modal Fake News Detection}
Multi-modal fake news detection combines texts and visuals to distinguish fake news. It is suggested that the rich visual information is helpful for fake news detection \cite{jin2017multimodal}. We collected image URLs of the news article and tweets, which could be used to develop multi-modal methods which aims to directly find clues in the visuals. Moreover, the user profile images could also reveal the credibility of the user's posts \cite{morris2012tweeting}. Therefore, we also collect the user profiles with a large number of profile image links. Future multi-modal models may take advantage of the images of news, tweets and users we obtained.   

\subsection{Fake Health News Early Detection}
The aim of fake health news early detection is to identify the fake news before it has been widely spread to users. One of the challenges is the limited social engagements  at the early stage of fake news propagation. Thus, obtaining various types of social engagements including tweets, retweets, replies is necessary. Besides, the user characteristics can play an important role for the early detection. However, many of the attempts of utilizing user information often simply make use of the profiles or the engagements with the news. In our datasets, every user interacting with the health news is supplied with his own profile, recent timelines and friend profiles. Overall, our datasets enable potential investigators extract more effective features of the users and social engagements for the fake health news early detection. 

\subsection{Fake Health News Propagation} 
In our analysis, we find the propagation of real and fake health news is complex.  Normal users and bots all contribute to the dissemination of both fake and real news on social media. The impact of health news on the social media platform differs a lot.
With the tweets, retweets and replies to the original news in the datasets, we can better understand the propagation of health news across health topics. What's more, the difference of spreading real and fake news could be explored. We could also discover the fake health news with high impact in the propagation analysis to figure out the most harmful fake health news.

\section{Conclusion}

In this paper, we release a comprehensive data repository FakeHealth, which containing two feature-rich health news datasets, to facilitate the research in fake health news domain. The data repository contains plenty news contents, massive social engagements, large user-user social networks and comprehensive explanations to the ground truth. We further conduct the exploratory analyses to show the the characteristics of the datasets. Through our analyses, we find the potential useful patterns and challenges in fake health news detection. Different types of fake news detection approaches are also evaluated on FakeHealth. The results provide baselines for further studies and demonstrate the quality of our datasets. The abundant information in FakeHealth make various novel and potential research directions possible.

\thanks
\section{Acknowledgements}
This material is based upon work supported by, or in part by, the National Science Foundation (NSF) under grant number \#IIS-1909702.

\appendix
\section{Appendix}

\setlength{\tabcolsep}{2pt}
\begin{table}[h]
    \small
    \centering
    \begin{tabular}{p{0.15\columnwidth}p{0.80\columnwidth}}
    \toprule
    Number & Criteria Questions\\
    \midrule
    C1 & Does it compare the new approach with existing alternatives?\\
    C2 & Does it adequately explain/quantify the harms of the intervention?\\
    C3 & Does it seem to grasp the quality of the evidence?\\
    C4 & Does it adequately quantify the benefits of the treatment/test/product/procedure?\\
    C5 & Does it establish the true novelty of the approach?\\
    C6 & Does it establish the availability of the treatment/test/product/procedure?\\
    C7 & Does it commit disease-mongering?\\
    C8 & Does it adequately discuss the costs of the intervention?\\
    \midrule
    S9 & Does the story use independent sources and identify conflicts of interest?\\
    S10 & Does the story appear to rely solely or largely on a news release?\\
    \midrule
    R9 & Does the news release identify funding sources \& disclose conflicts of interest?\\
    R10 & Does the news release include unjustifiable, sensational language, including in the quotes of researchers?\\
    \bottomrule
    \end{tabular}
    \caption{The criteria in the news reviews for news story and news release. C1-C8 denote the eight common criteria. S9 and S10 are criteria for the news story. R9 and R10 represents two special criteria for the news release. }
    \label{tab:criteria}
    \vspace{-3mm}
\end{table}

\subsection{Data Format}
FakeHealth consists of four categories of information, i.e., news contents, news reviews, social engagements and user network. They are stored in four folders. 
\begin{itemize}
    \item \textit{contents} has two subfolders: \textit{HealthStory} and \textit{HealthRelease}. Each folder lists the josn files of the news contents, which is named as the news ID. 
    \item \textit{reviews} includes two json files: \textit{HealthStory.json} and \textit{HealthRelease.json}. These json files contain the news reviews for both datasets.
    \item \textit{engagements} includes two folders: \textit{HealthStory} and \textit{HealthRelease}. Tweets, retweets and replies are listed in separate folders here. Each engagement is stored in a json file in the name of its ID.   
    \item \textit{user\_network} includes four folders to store user profiles, user timelines, user followers and user followings. All the users in FakeHealth which includes HealthRelease and HealthStory are covered here. The profiles, timelines, follower profiles and following profiles are saved as json files named in the ID of the corresponding user.
\end{itemize} 
Due to the Twitter policy of protecting user privacy, the full contents of user social engagements and network are not allowed to directly publish. Instead, we store the IDs of all social engagements of HealthStory and HealthRelease into two json files. More specially, each json file contains a dictionary whose key is the the news ID and the value contains the IDs of tweets, replies and retweets. With these two files and the API we provided for FakeHealth, researchers can trivially obtain the full contents of social engagements and user network from Twitter. We will also maintain and update the repository to ensure its usability.  
\newpage

\bibliographystyle{aaai}
\bibliography{Bibliography}
\end{document}